\documentclass[aps,nofootinbib,noshowpacs,showkeys,preprintnumbers,amsmath,amssymb]{revtex4}
\usepackage{graphicx,color}
\usepackage{graphics}
\usepackage{rotating}
\usepackage{amssymb}

\usepackage{color}
\usepackage{epsfig}
\usepackage{dcolumn}% Align table columns on decimal point
\usepackage{bm}% bold math
\usepackage{multirow}

\def\be{\begin{equation}}
\def\ee{\end{equation}}
\def\ba{\begin{eqnarray}}
\def\ea{\end{eqnarray}}
\def\bea{\begin{eqnarray}}
\def\eea{\end{eqnarray}}
\def\bes{\begin{subequations}}
\def\ees{\end{subequations}}
\def\bear{\begin{array}}
\def\eear{\end{array}}

\usepackage{epstopdf}

\newcommand{\A}{{\mathcal{A}}}
\newcommand{\tA}{{\widetilde {\mathcal{A}}}}
\newcommand{\ta}{{\widetilde a}}

\newcommand{\MSbar}{\overline{\rm MS}}

\begin{document}
\preprint{USM-TH-353}

\title{Bjorken sum rule in QCD frameworks with analytic (holomorphic) coupling}
\author{C\'esar Ayala$^{1}$}
\author{Gorazd Cveti\v{c}$^1$} 
\author{Anatoly V.~Kotikov$^2$}
\author{Binur G.~Shaikhatdenov$^2$}
\affiliation{$^1$Department of Physics, Universidad T{\'e}cnica Federico
Santa Mar{\'\i}a (UTFSM), Casilla 110-V, Valpara{\'\i}so, Chile\\
$^2$Joint Institute for Nuclear Research, $141980$ Dubna, Russia}

\date{\today}

\begin{abstract}
We investigate the Bjorken polarized sum rule (BSR) in three approaches to QCD with analytic (holomorphic) coupling: Analytic Perturbation Theory (APT), Two-delta analytic QCD (2$\delta$anQCD), and Three-delta lattice-motivated analytic QCD in the three-loop and four-loop MOM scheme (3l3$\delta$anQCD, 4l3$\delta$anQCD). These couplings do not have unphysical (Landau) singularities, and have finite values when the transferred momentum goes to zero, which allows us to explore the infrared regime. With the exception of APT, these theories at high momenta practically coincide with the underlying perturbative QCD (pQCD) in the same scheme. We apply them in order to verify the Bjorken sum rule within the range of energies available in the data collected by the experimental JLAB collaboration, i.e., $0.05 \ {\rm GeV}^2 <Q^2< 3 \ {\rm GeV}^2$ and compare the results with those obtained by using the perturbative QCD coupling. The results of the new frameworks with respective couplings (2$\delta$ and 3$\delta$) are in good agreement with the experimental data for $0.5 \ {\rm GeV}^2<Q^2<3 \ {\rm GeV}^2$ already when only one higher-twist term is used. In the low-$Q^2$ regime ($Q^2 \lesssim  1 \ {\rm GeV}^2$) we use $\chi$PT-motivated expression \textcolor{black}{or an expression motivated by the light-front holography (LFH) QCD} used earlier in the literature.
\end{abstract}
\pacs{11.55.Hx, 11.55.Fv, 12.38.Bx, 12.38.Lg}
\keywords{perturbation expansion in low-energy QCD; IR-safe QCD coupling; holomorphic behavior; spacelike quantities; QCD phenomenology}
\maketitle

\section{Introduction}
\label{sec:intr}

The spin structure in deep inelastic scattering (DIS) is one of the biggest 
challenges of hadronic physics due to its nonperturbative (NP) nature. The
analysis of such observable can be performed with the help of operator product 
expansion (OPE) in conjunction with the underlying perturbative QCD (pQCD). 
The OPE formalism is important and an essential ingredient to explore 
DIS at moderately low energies $Q^2\sim 1\text{GeV}^2$ ($Q^2=-q^2$ is the 
momentum transfer of the process in the Euclidean domain).

A good candidate to test the nonperturbative behavior in QCD is the well known Bjorken polarized sum rule (BSR) $\Gamma_1^{p-n}$ \cite{BjorkenSR}, for which the recent experimental data given by Jefferson Lab (JLAB) is avaliable in the range $0.05 \ {\rm GeV}^2 < Q^2 < 3 \ {\rm GeV}^2$ \cite{data2,data2B,data2N} (and even beyond), as well as those by SLAC \cite{data3}, and the theoretical perturbation expansion of the leading-twist (LT) contribution to BSR is now known to ${\rm N}^3{\rm LO}$ ($\sim \alpha_s^4$) \cite{nnnloBSR}. We will investigate the applicability of pQCD together with one higher-twist (HT) contribution $\sim 1/Q^2$ dictated by OPE. At low momenta $Q^2 < Q_0^2$ ($\approx 0.4$-$0.6 \ {\rm GeV}^2$), we use a $\chi$PT-motivated expression \cite{data2B} with its first term ($\sim Q^2$) fixed by the Gerasimov-Drell-Hearn sum rule \cite{GDHsr}, \textcolor{black}{or an expression motivated by the light-front holography (LFH) QCD \cite{LFH}.}
%We pay a particular attention to this aspect, since at very low energies ($Q^2\lesssim0.3\text{GeV}^2$) the HT blows up and then we propose a phenomenological modification, where the new $Q^2$-dependence of HT is acquired by the simple change $Q^2\mapsto Q^2+M^2$, with $M^2$ an infrared safety mass parameter (SIRM).

In addition to pQCD, we will consider, for evaluation of BSR, analytic frameworks of QCD (anQCD) which provide us with a useful  
tool to evaluate physical quantities at low-momentum transfer. In the anQCD frameworks the running coupling
has no spurious (Landau) singularities, unlike perturbative QCD (pQCD) in the usual schemes such as $\MSbar$. Such pQCD has a coupling $a_{\rm pt}(Q^2) \equiv \alpha_s(Q^2)/\pi$ which, for the general complex spacelike momenta $Q^2 \in \mathbb{C} \backslash (-\infty, 0]$, has Landau singularities  at small momenta $|Q^2| \lesssim 1 \ {\rm GeV}^2$, while the general principles of quantum field theories (QFT) dictate that the spacelike QCD observables ${\cal D}(Q^2)$, such as current correlators and structure functions, are holomorphic (analytic) functions of $Q^2$ in the entire
generalized spacelike region $Q^2 \in \mathbb{C} \backslash (-\infty, 0]$.
Since the usual pQCD couplings $a_{\rm pt}(Q^2)$ do not reflect these properties, any evaluation
of (the LT part of) 
${\cal D}(Q^2)$ in terms of $a_{\rm pt}(Q^2)$ does not reflect these properties dictated by QFT.
On the other hand, in anQCD we have $a_{\rm pt}(Q^2) \mapsto \A(Q^2)$, where
$\A(Q^2)$ is the anQCD coupling holomorphic in 
$Q^2 \in \mathbb{C} \backslash (-\infty, 0]$. 
As a consequence, the evaluation of the LT contribution 
${\cal D}_{\rm eval.}(Q^2) \mapsto  {\cal F}(\A(k Q^2))$
has the correct analyticity properties (as is also the case of the
HT contribution). Here, $k \sim 1$ is the renormalization scale
parameter.

We will consider three different anQCD frameworks.  
The first one is the Analytic Perturbation Theory (APT) of
Shirkov, Solovtsov {\it et al.\/}~\cite{ShS,MS96,ShS98,Sh}.
In this framework, the discontinuity function
$\rho_1^{\rm (pt)}(\sigma) \equiv {\rm Im} a_{\rm pt}(Q^2=-\sigma - i \epsilon)$
of the underlying QCD
is kept unchanged on the entire negative axis in the $Q^2$-plane,
and is zero on the Landau cut of $a_{\rm pt}(Q^2)$
(at $-\Lambda_{\rm Lan.}^2 \leq \sigma < 0$):
${\rm Im} \A^{\rm (APT)}(-\sigma - i \epsilon) =
\rho_1^{\rm (pt)}(\sigma)$ for $\sigma >0$, and
${\rm Im} \A^{\rm (APT)}(-\sigma - i \epsilon) = 0$ for $\sigma < 0$.
The APT coupling $\A^{\rm (APT)}(Q^2)$ 
for $Q^2 \in \mathbb{C} \backslash (-\infty, 0]$ is then arrived at by
using a dispersion relation involving 
${\rm Im} \A^{\rm (APT)}(-\sigma - i \epsilon)$.
The APT-analogs of the integer powers $a_{\rm pt}(Q^2)^n$,
$\A_n^{\rm (APT)}(Q^2)$,
were also obtained in these works.
The extension to the analogs $\A_{\nu}^{\rm (APT)}(Q^2)$
of powers $a_{\rm pt}(Q^2)^{\nu}$ for noninteger $\nu$ in this framework was performed
and applied in the works ~\cite{BMS05,BKS05,BMS06,BMS10} known as Fractional APT (FAPT).

The other considered analytic frameworks are the Two-delta analytic QCD (2$\delta$anQCD, \cite{2danQCD}) and the lattice-motivated Three-delta analytic QCD (3$\delta$anQCD, \cite{3l3danQCD,4l3danQCD}). In their construction, they are less closely than APT based on the underlying pQCD coupling $a_{\rm pt}(Q^2)$: the equality ${\rm Im} \A(-\sigma - i \epsilon) = \rho_1^{\rm (pt)}(\sigma)$ is taken only for sufficiently large $\sigma \geq M_0^2$  (where $M_0 \sim 1$ GeV is ``pQCD-onset'' scale). For low positive $\sigma$, $0 < \sigma < M_0^2$, the otherwise unknown discontinuity function $\rho_1(\sigma) \equiv {\rm Im} \A(Q^2=-\sigma - i \epsilon)$ is parametrized by two or three delta functions. Such an ansatz is partly motivated by the Pad\'e approximant approach to the coupling $\A(Q^2)$.  The $\Lambda$ scale parameter ($\sim 0.1$-$1$ GeV) of the underlying pQCD is determined by the high-energy QCD phenomenology, i.e., by the world average value of $\alpha_s(M_Z^2)$ in the scheme of the underlying pQCD (and thus, indirectly, by the world average value in $\MSbar$ scheme). The other parameters of the framework, namely those of the delta functions and the scale $M_0$, are then constrained by requiring that the framework agree to a high degree of precision with the underlying pQCD for large $|Q^2| >  \Lambda^2$, and by an additional requirement that the framework reproduce the experimentally well measured value $r_{\tau} \approx 0.20$ of the $\tau$ lepton semihadronic nonstrange $V+A$ decay rate ratio (i.e., a well measured $\sim 1$ GeV QCD quantity with suppressed HT contribution). The renormalization schemes of the underlying pQCD coupling in 2$\delta$anQCD are restricted by requiring acceptable values of $M_0 \sim 1$ GeV and $\A(0) \sim 1$ \cite{2danQCD,anOPE,mathprg}. In lattice-motivated 3$\delta$anQCD, the renormalization scheme is taken to be the MiniMOM scheme \cite{MiniMOM,BoucaudMM,CheRet} used in the high-volume lattice calculations of the lattice coupling \cite{LattcoupNf0,LattcoupNf0b,LattcoupNf2} via the calculation of the low-momentum gluon and ghost dressing functions in the Landau gauge. In Ref.~\cite{3l3danQCD}, the scheme for the underlying pQCD coupling is taken to be ($N_f=3$) MiniMOM at the three-loop level, and in Ref.~\cite{4l3danQCD} at the four-loop level. The lattice calculations provide two conditions for the coupling in 3$\delta$anQCD at very low $Q^2 < 1 \ {\rm GeV}^2$, which give additional constraints on the delta functions.

The construction of analytic analogs $\A_{n}(Q^2)$ of the powers $a_{\rm pt}(Q^2)^{n}$, for general anQCD, was formulated in Ref.~\cite{Cvetic:2006gc} for $n$ integer and in Ref.~\cite{GCAK} for general (noninteger) $n$.

For $Q^2 > Q_0^2$ ($\approx 0.5 \ {\rm GeV}^2$), the recent four-loop pQCD series of the LT BSR function \cite{nnnloBSR} will be used for an accurate and updated analysis with pQCD and of anQCD frameworks; we will include in the analysis one HT term $\mu_4^{p-n}/Q^2$. At $Q^2 < Q_0^2$, a $\chi$PT-motivated expression \textcolor{black}{or a LFH QCD-motivated expression} for the BSR function will be used. The implementation of the aforementioned frameworks will be given by a fit to the experimental data, of the free NP parameters $\mu_4^{p-n}$, $Q_0^2$ and $A$ (the latter of the $\chi$PT form), and of the renormalization scale (RScl) parameter $C \equiv \ln(\mu^2/Q^2)$ appearing in the LT contribution. At the high-low border point $Q^2=Q_0^2$ the OPE expression (LT+HT) and the $\chi$PT-motivated expressions will be required to coincide (matching condition).

The paper is organized as follows. In Sec.~\ref{sec:models} we provide a brief description of the three mentioned analytic QCD frameworks that we will use in the evaluation of the BSR function $\Gamma_1^{p-n}(Q^2)$: (F)APT, 2$\delta$anQCD, and  3$\delta$anQCD. We also describe the implementation of holomorphic (analytic) nonpower series from the usual perturbation series. In Sec.~\ref{sec:BSR} we present the theoretical basis, with expressions for LT and HT contributions of the BSR function, and the analytized version of the LT contribution. We analyze the convergence of the new analytic series for each anQCD framework. We describe \textcolor{black}{for $Q^2 \leq Q_0^2$ the $\chi$PT-motivated  and the LFH QCD-motivated expressions, and for $Q^2 \geq Q_0^2$} the OPE expression, to be used for the BSR function. In Sec.~\ref{sec:num} we fit the BSR function with the combined JLAB and SLAC data for the mentioned anQCD frameworks \textcolor{black}{and discuss the obtained results}. In Sec.~\ref{sec:concl} we summarize our results. In Appendix \ref{App:A} we present the conventions and formulas of the underlying pQCD coupling and their beta function, \textcolor{black}{and in Appendix \ref{App:B} we discuss the charm mass contributions.}

\section{Analytic QCD frameworks: (F)APT, 2$\delta$anQCD and 3$\delta$anQCD}
\label{sec:models}

In analytic frameworks of QCD the running coupling $\A(Q^2)$, corresponding to $a_{\rm pt}(Q^2) \equiv \alpha_s(Q^2)/\pi$ in pQCD, has the analytic (holomorphic) properties
in the complex $Q^2$-plane which are qualitatively equal to the analytic
properties of the spacelike observables ${\cal D}(Q^2)$ such
as current correlators and structure functions. Namely, 
the general principles of quantum field theory (locality, unitarity,
microcausality) dictate that ${\cal D}(Q^2)$ is an analytic
(holomorphic) function of complex $Q^2$ in the entire
complex plane with the possible exception (on parts) of the
negative semiaxis \cite{BS,Oehme}. More specifically, ${\cal D}(Q^2)$ is
holomorphic for $Q^2 \in \mathbb{C} \backslash (-\infty, 0]$,
or, for $Q^2 \in \mathbb{C} \backslash (-\infty, -M_{\rm thr}^2]$
where $M_{\rm thr} \sim 0.1$ GeV is a threshold scale. 
The coupling $\A(Q^2)$ in analytic QCD is required to fulfill the
same analyticity condition. This requirement is motivated by the
fact that (the leading-twist part of) ${\cal D}(Q^2)$ can be evaluated
as a function of $\A(k Q^2)$ where $k \sim 1$ is the renormalization
scale (RScl) parameter: ${\cal F}(\A(k Q^2))$.
Such a holomorphic running coupling $\A(Q^2)$ 
can be written by the use of Cauchy theorem as a dispersion integral along
the cut $\sigma \equiv -Q^{' 2} \in [M_{\rm thr}^2,\infty)$
\be
\A(Q^2) = \frac{1}{\pi} \int_{\sigma=M^2_{\rm thr}}^{\infty} \frac{d \sigma \rho_1(\sigma)}{(\sigma + Q^2)} \ ,
\label{Adisp}
\ee
where the discontinuity function $\rho_1(\sigma) \equiv 
{\rm Im} \A(-\sigma - i \varepsilon)$ 
is nonzero for $\sigma \geq M_{\rm thr}^2$.
In Eq.~(\ref{Adisp}), $Q^2$ can be any value in the 
complex plane except on the cut $(-\infty, -M^2_{\rm thr}]$. 
We note that in the usual pQCD case $M^2_{\rm thr}=-\Lambda_{\rm QCD}^2-0$,
i.e., the cut extends to the ``Landau'' region 
$-\Lambda_{\rm QCD}^2 \leq \sigma < 0$
($0 < Q^{'2} \leq \Lambda_{\rm QCD}^2$), where 
$\Lambda_{\rm QCD}^2 \sim 0.1 \ {\rm GeV}^2$.

Due to asymptotic freedom, the perturbative structure prevails at large 
$\sigma \gg \Lambda_{\rm QCD}^2$ where $\rho_1(\sigma) \approx
\rho_1^{\rm (pt)}(\sigma)$.

In general, anQCD frameworks have $\A(Q^2)$ which differs from the
underlying pQCD running coupling $a_{\rm pt}(Q^2)$ by nonperturbative
(nonanalytic in $a_{\rm pt}$) terms $\sim (\Lambda^2/Q^2)^N$:
\be
\A(Q^2) - a_{\rm pt}(Q^2) \sim \left( \frac{\Lambda^2}{Q^2} \right)^N
\qquad {\rm for} \; |Q^2| > \Lambda^2 \sim 0.1 \ {\rm GeV}^2,
\label{diff}
\ee
where $N \geq 1$ is a positive integer.
In such anQCD frameworks, the analytic analogs $\A_n$ of pQCD powers
$a_{\rm pt}^n$ are not powers $\A^n$: $((a_{\rm pt})^n)_{\rm an.} \not=
((a_{\rm pt})_{\rm an.})^n$, i.e., $\A_n \not= \A^n$. 
It turns out that, for integer values of $n$, 
it is convenient to analytize first 
the logarithmic derivatives $\ta_{{\rm pt},n}(Q^2)$ 
which, in contrast to powers, are ``respected'' under the
analytization, $\ta_{{\rm pt},n}(Q^2) \mapsto \tA_n(Q^2)$
\cite{Cvetic:2006gc}
\be
\ta_{{\rm pt},n+1}(Q^2) \equiv \frac{(-1)^n}{\beta_0^n n!}
\frac{ \partial^n a_{\rm pt}(Q^2)}{\partial (\ln Q^2)^n} \ ,
\qquad
\tA_{n+1}(Q^2) \equiv \frac{(-1)^n}{\beta_0^n n!}
\frac{ \partial^n \A(Q^2)}{\partial (\ln Q^2)^n} \ ,
\qquad (n=0,1,2,\ldots) \ .
\label{tAn}
\ee
Namely, we have $(\ta_{{\rm pt},n+1})_{\rm an} = \tA_{n+1}$. The power analogs
$\A_n$ are then linear combinations of logarithmic derivatives
${\widetilde \A}_{n+m}$
\be
\A_n = {\widetilde \A}_n + \sum_{m=1}^{\infty} {\widetilde k}_m(n) {\widetilde \A}_{n+m},
\label{AntAn}
\ee
where the coefficients ${\widetilde k}_m(n)$ were obtained in  
Ref.~\cite{Cvetic:2006gc} for integer $n$,
and in Ref.~\cite{GCAK} for general noninteger $n$.

\subsection{(Fractional) Analytic Perturbation Theory ((F)APT)}
\label{subs:FAPT}

The first explicitly constructed analytic QCD framework in the literature
is the well-known Analytic Perturbation Theory (APT) 
of Shirkov, Solovtsov and Milton \cite{ShS,MS96}. It is constructed 
from pQCD, when in the dispersion integral (\ref{Adisp}) the
discontinuity function is kept unchanged,
$\rho_1(\sigma) = \rho_1^{\rm (pt)}(\sigma)$
[where: $\rho_1^{\rm (pt)}(\sigma) \equiv 
{\rm Im} a_{\rm pt}(-\sigma - i \epsilon)$],
and the cut in the Landau region is removed
[$\rho_1^{\rm (pt)}(\sigma) \mapsto 0$ for 
$ - \Lambda_{\rm Lan.} \leq \sigma < 0$]
\be
\A^{\rm {(APT)}}(Q^2)
%\equiv\left(a(Q^2) \right)_{an} 
= \frac{1}{\pi} \int_{\sigma= 0}^{\infty}
\frac{d \sigma \rho_1^{\rm (pt)}(\sigma) }{(\sigma + Q^2)} \ .
\label{MAA1disp}
\ee
In APT, the analogs of general powers of the running coupling 
$a^\nu (Q^2)$ (with $\nu$ general real) can be constructed in the specific
APT-way: as in Eq.~(\ref{MAA1disp}), 
but replacing $\rho_1^{\rm (pt)}(\sigma) \mapsto
\rho_{\nu}^{\rm (pt)}(\sigma) \equiv 
{\rm Im} a_{\rm pt}^{\nu}(Q^{'2}=-\sigma - i \epsilon)$ \cite{MS96,Sh}
\begin{equation}
{\A}^{\rm {(APT)}}_{\nu}(Q^2)\equiv\left(a^\nu(Q^2) \right)_{an}^{\rm {(APT)}} 
= \frac{1}{\pi} \int_{\sigma= 0}^{\infty}
\frac{d \sigma {\rho^{\rm {(pt)}}_{\nu}}(\sigma) }{(\sigma + Q^2)} \ .
\label{AnuMA}
\end{equation}
Refs.~\cite{BMS05,BKS05,BMS06,BMS10} obtained and applied the
explicit expressions for $\A^{\rm {(APT)}}_{\nu}$ at one-loop
level of the underlying pQCD, and extensions at higher-loop
level.
This theory is usually named Fractional APT (FAPT);
cf.~Refs.~\cite{Bakulev} for reviews of FAPT.

The one-loop (LO) APT coupling ($\nu=1$ in (F)APT) has the form
\be 
\A^{\rm {(APT),LO}}(Q^2)=\frac{1}{\beta_0}\left(\frac{1}{{\rm ln}
(Q^2/\Lambda_{{\rm LO}}^2)}+\frac{\Lambda_{{\rm LO}}^2}{\Lambda_{{\rm LO}}^2-Q^2}\right)
= a^{\rm LO}(Q^2) - \frac{1}{\beta_0}
 \frac{\Lambda^2_{\rm LO}}{Q^2 - \Lambda^2_{\rm LO}} \ .
\ee
The corresponding generalization to one-loop FAPT, derived
and applied in Ref.~\cite{BMS05}, is
\be
\A^{\rm {(FAPT)LO}}_{\nu}(Q^2) = \frac{1}{\beta_0^{\nu}}
\left(  \frac{1}{\ln^{\nu}(1/z_{\rm LO})} -
\frac{ {\rm Li}_{1-\nu}(z_{\rm LO})}{\Gamma(\nu)} \right)=
a_{\rm LO}(Q^2)^{\nu} - \frac{1}{\beta_0^{\nu}}
 \frac{\rm{Li}_{1-\nu}(z_{\rm LO})}{\Gamma(\nu)},
\label{MAAnu1l}
\ee
where, $z_{\rm LO} \equiv \Lambda_{\rm LO}^2/Q^2$ and 
${\rm Li}_{1-\nu}(z)$ is the polylogarithm function of order $1-\nu$.

At two-loop (NLO) level, there are no exact results and we can 
use in the above general formula (\ref{AnuMA})
the discontinuities ${\rho_{\nu}^{\rm {(pt)}}}(\sigma)$
obtained from the underlying exact two-loop
running coupling $a_{\rm pt}(Q^2)$ which contains Lambert function
$W_{\pm 1}(z)$, cf.~Eq.~(\ref{aptexact}). 
In Refs.~\cite{BK1,BK2,mathprg,mathprgb}
numerical algorithms were performed (in \cite{BK1,mathprgb} in Maple and/or Fortran; in \cite{BK2,mathprg} in Mathematica) for calculation of any 
$A_{\nu}^{\rm (FAPT)}$ up to four-loop level, i.e., using for $a_{\rm pt}$
the four-loop ($\MSbar$) coupling.

\subsection{Two-delta anQCD framework (2$\delta$anQCD)}
\label{subs:2dan}

Another analytic QCD framework we will use here is 2$\delta$anQCD 
\cite{2danQCD,anOPE}. Here,
the discontinuity function ${\rho_1^{(2\delta)}}(\sigma)$ is approximated at high 
scales $\sigma \geq M_0^2$ ($\sim \ 1 \ {\rm GeV}^2$) by 
${\rho_1^{\rm {(pt)}}}(\sigma) \equiv {\rm Im} a_{\rm pt}(- \sigma - i \epsilon)$ 
of the underlying pQCD. 
In the unknown low-scale regime $0<\sigma<M_0^2$, 
${\rho_1^{(2\delta)}}(\sigma)$ is parametrized by two delta functions
\be
{\rho_1^{(2\delta)}}(\sigma) = \pi F_1^2 \delta(\sigma-M_1^2) +
\pi F_2^2 \delta(\sigma-M_2^2) + \Theta(\sigma-M_0^2){\rho_1^{\rm {(pt)}}}(\sigma)
\ ,
\label{rhoA2d}
\ee
where $0 < M_1 < M_2 < M_0 \sim 1$ GeV.
The dispersion relation (\ref{Adisp}) with  ${\rho_1^{(2\delta)}}(\sigma)$
gives the following coupling:
\be
\A^{\rm {(2\delta)}}(Q^2)
%\equiv\left(a(Q^2) \right)_{an} = 
= \frac{F_1^2}{Q^2+M_1^2} + \frac{F_2^2}{Q^2+M_2^2} +
\frac{1}{\pi} \int_{\sigma= M_0^2}^{\infty}
\frac{d \sigma {\rho_1^{\rm {(pt)}}}(\sigma) }{(\sigma + Q^2)} \ .
\label{2dA1disp}
\ee
We use the Lambert-scheme coupling Eq.~(\ref{aptexactc2}) 
for the underlying pQCD coupling, with $c_2 = \beta_2/\beta_0$ 
a chosen scheme parameter.
The 2$\delta$anQCD framework uses information on the underlying pQCD coupling $a_{\rm pt}$ and thus on the pQCD discontinuity function $\rho_1^{\rm (pt)}(\sigma)$. This quantity is fixed by the choice of the Lambert scale parameter $\Lambda_{\rm L.}$ of Eq.~(\ref{zexpr}), or equivalently, by the choice of the (world average) value of $\alpha_s(M_Z^2;\MSbar)$. The other five parameters are explicitly visible in Eqs.~(\ref{rhoA2d})-(\ref{2dA1disp}):
$M_j^2 = s_j \Lambda_{\rm L.}^2$ and $F_j^2= f_j^2 \Lambda_{\rm L.}^2$ ($j=1,2$),
and $M_0^2= s_0 \Lambda_{\rm L.}^2$.
These five parameters get their values fixed by 
altogether five conditions, namely: (1) the condition that at high $|Q^2|$ the coupling $\A(Q^2)$ practically coincides with the underlying pQCD coupling, Eq.~(\ref{diff}), with $N=5$ (these are four conditions, because in general $N=1$), which determines $s_1$, $s_2$, $f_1^2$ and $f_2^2$ as functions of $s_0$; (2) the condition that the obtained 2$\delta$anQCD reproduces the correct value of the semihadronic strangeless $\tau$ lepton decay ratio $r_{\tau}$ of the $V+A$ channel (i.e., the most precisely measured $\sim 1$ GeV QCD quantity with strongly suppressed HT contribution), which determines the parameter $s_0$ (pQCD-onset parameter). The parameter $c_2 \equiv \beta_2/\beta_0$ of the scheme of the underlying pQCD is chosen in a preferred interval, $-5.6 < c_2 < -2$, with the preferred central value $c_2=-4.9$. We refer for conventions to Appendix \ref{App:A}. For details of the framework, we refer to \cite{2danQCD,anOPE,mathprg}, and in particular to Table 2 of Ref.~\cite{mathprg}. Evaluations of $\A_n$, analogs of the powers $a_{\rm pt}^n$, are performed with the help of logarithmic derivatives, via Eqs.~(\ref{tAn})-(\ref{AntAn}). In Refs.~\cite{2danQCD,anOPE,mathprg} the parameters of the framework were adjusted so that the underlying coupling corresponded to the $\MSbar$ value $\alpha_s(M_Z^2;\MSbar)=0.1184$ and the $D=0$ semihadronic tau decay ratio $r_{\tau}$ (canonical, strangeless and massless) was equal to $r_{\tau}^{(D=0)}=0.203$ in the leading-$\beta_0$ (LB) plus beyond-LB (bLB) approach. Here we will keep the value $c_2=-4.9$ as in \cite{mathprg}, but will take as the reference value $\alpha_s(M_Z^2;\MSbar)=0.1185$ \cite{PDG2014} and  $r_{\tau}^{(D=0)}=0.201$ (in LB+bLB approach), in order to keep these values equal to those taken in the (lattice-motivated) 3$\delta$anQCD \cite{3l3danQCD,4l3danQCD}. The resulting parameters are given in Table \ref{tab2dan}.
\begin{table}
\caption{Values of the parameters of 2$\delta$anQCD, for $N_f=3$ and $c_2=-4.9$ in the Lambert scheme: the Lambert $\Lambda_{{\rm L.}}$ scale; $s_j=M_j^2/\Lambda_{{\rm L.}}^2$ ($j=0,2,1$) and $f_j^2=F_j^2/\Lambda_{{\rm L.}}^2$, for our choice of $\alpha_s(M_Z^2;\MSbar)=0.1185$ and $r_{\tau}^{(D=0)}=0.201$. For comparison, also the values for the choice $\alpha_s(M_Z^2;\MSbar)=0.1184$ and $r_{\tau}^{(D=0)}=0.203$ are given (cf.~\cite{mathprg} Table 2, third line there).}
\label{tab2dan}
\begin{ruledtabular}
\centering
\begin{tabular}{rr|llllllll}
%\hline
$\alpha_s(M_Z^2;\MSbar)$ & $r_{\tau}^{(D=0)}$ & $\Lambda_{{\rm L.}}$ [GeV] & $s_0$ & $s_1$ & $f_1^2$ & $s_2$ & $f_2^2$ & $M_0$ [GeV] & $\A(0)$
\\
\hline
0.1185 & 0.201 & 0.2564 & 25.610 &  18.734 & 0.2929 & 1.0361 & 0.5747 & 1.298 & 0.6593 \\
\hline
0.1184 & 0.203 & 0.2552 & 23.076 &  16.839 & 0.2746 & 0.7688 & 0.5505 & 1.226 & 0.8231 \\
\end{tabular}
\end{ruledtabular}
\end{table}

\subsection{Three-delta lattice-motivated anQCD (3$\delta$anQCD)}
\label{subs:3dan}

The ansatz is very similar to the one for 2$\delta$anQCD, but now the low-$\sigma$ region is parametrized by three delta functions and not all of them have positive coefficients 
\be
\rho_1(\sigma) =  \pi \sum_{j=1}^{3} {\cal F}_j \; \delta(\sigma - M_j^2)  + \Theta(\sigma - M_0^2) \rho_1^{\rm (pt)}(\sigma) \ ,
\label{rhoA3d}
\ee
Consequently, the considered coupling is
\bea
\A^{(3 \delta)}(Q^2) & = &  \sum_{j=1}^3 \frac{{\cal F}_j}{(Q^2 + M_j^2)} + \frac{1}{\pi} \int_{M_0^2}^{\infty} d \sigma \frac{ \rho_1^{\rm (pt)}(\sigma) }{(Q^2 + \sigma)} \ .
\label{AQ2}
\eea
The underlying pQCD coupling $a_{\rm pt}$, and thus the discontinuity function $\rho_1^{\rm (pt)}(\sigma)$, are fixed again by a chosen reference value (world average) of $\alpha_s(M_Z^2;\MSbar)$. The other seven parameters, namely ${\cal F}_j$, $M_j^2$ ($j=1,2,3$) and $M_0^2$, are fixed by a total of seven conditions at high momentum, intermediate momentum, and low momentum:  (1) the condition that the coupling  $\A(Q^2)$ at $|Q^2| > 1 \ {\rm GeV}^2$ practically coincides with the underlying pQCD coupling, Eq.~(\ref{diff}), with $N=5$ (this represents again four conditions); (2) the condition that the obtained 3$\delta$anQCD reproduces at $|Q^2| \sim 1 \ {\rm GeV}^2$ the correct value of the semihadronic strangeless $\tau$ lepton decay ratio $r_{\tau}$ of the $V+A$ channel; (3) the condition that for positive $Q^2$ the coupling $\A(Q^2)$ has local maximum at $Q^2 \approx 0.13$-$0.14 \ {\rm GeV}^2$; (4) the condition that at $|Q^2| < 0.1 \ {\rm GeV}^2$ the coupling behaves as $\A(Q^2) \sim Q^2$ when $Q^2 \to 0$. The two conditions (3) and (4) are lattice-motivated, because high-volume lattice calculations of the gluon and ghost dressing functions in the Landau gauge suggest such behavior \cite{LattcoupNf0} (cf.~also \cite{LattcoupNf0b}). These lattice calculations were performed for $N_f=0$; similar results are obtained when $N_f=2$ \cite{LattcoupNf2}, although there the precision of lattice calculation is not so high. We will take $N_f=3$ in our coupling, in order to have a coupling applicable reasonably well to the entire region $|Q^2| < 10 \ {\rm GeV}^2$. The mentioned lattice calculations were performed in the MiniMOM (MM) renormalization scheme \cite{MiniMOM,BoucaudMM,CheRet}, and we constructed our coupling with the underlying pQCD coupling in the scheme which either agrees with the $N_f=3$ MiniMOM scheme at the three-loop level (3l3$\delta$anQCD, \cite{3l3danQCD}) or at the four-loop level (4l3$\delta$anQCD, \cite{4l3danQCD}). Furthermore, we rescaled all the momentum scales from the lattice MiniMOM convention ($\Lambda_{\rm MM}$) to the usual scale convention ($\Lambda_{\MSbar}$), \textcolor{black}{this representing the so called Lambert MiniMOM scheme (LMM).} We note that the maximum of the lattice coupling $\A_{\rm latt.}(Q^2)$ is at about $0.45 \ {\rm GeV}^2$ in the lattice MiniMOM scale convention. For further details on the framework, we refer to Refs.~\cite{3l3danQCD,4l3danQCD} for the 3l3$\delta$anQCD and 4l3$\delta$anQCD cases.

Evaluations of $\A_n$, analogs of the powers $a_{\rm pt}^n$, in 2$\delta$anQCD and 3$\delta$anQCD are performed, as can be in any analytic QCD framework (even in (F)APT), with the help of logarithmic derivatives, via Eqs.~(\ref{tAn})-(\ref{AntAn}), Ref.~\cite{Cvetic:2006gc} for integer $n$, and Ref.~\cite{GCAK} for general (noninteger) $n$.

\section{Bjorken sum rule in Analytic QCD frameworks}
\label{sec:BSR}

\subsection{Perturbation expansion}
\label{subs:PT}

The polarized Bjorken sum rule (BSR) is defined as the nonsinglet combination given by the difference between proton and neutron polarized structure functions integrated over the whole $x$-Bjorken interval. It is represented by the BSR function $\Gamma_1^{p-n}$: 
\be
\Gamma_1^{p-n}(Q^2)=\int_0^1 dx \left[g_1^p(x,Q^2)-g_1^n(x,Q^2) \right]\ ,
\label{BSRdef}
\ee 
which is the first moment of the nonsinglet contribution to the polarized structure functions.

BSR can be written in terms of a sum of two series, one coming from pQCD as an expansion of the running coupling $a_{\rm pt}(Q^2)=\alpha_s(Q^2)/\pi$ and other from the higher-twist (HT) contributions dictated by the 
OPE \cite{BjorkenSR}
\be
\label{BSR}
\Gamma_1^{p-n}(Q^2)=\frac{g_A}{6}E_{\rm {NS}}(Q^2)+\sum_{i=2}^\infty
\frac{\mu_{2i}^{p-n}(Q^2)}{Q^{2i-2}}\ ,
\ee
 where in the limit when $Q^2\mapsto\infty$, $\Gamma_1^{p-n}(\infty)=g_A/6$, with the nucleon axial charge equal to $g_A=1.2723\pm0.0023$ \cite{PDG2016} (we will use the central value). In our notations, the $Q^2$-dependence will be implied in the HT coefficients, $\mu_{2i}^{p-n}\equiv\mu_{2i}^{p-n}(Q^2)$. We will include only the first HT term $\sim \mu_4^{p-n}$, \textcolor{black}{considering that this term may contain also compensations for higher order terms $\mu_{2 i}^{p-n}/(Q^2)^{2 i -2}$ ($i > 2$).}\footnote{\textcolor{black}{We note that often the explicit contributions of the higher terms $\mu_{2 i}^{p-n}/(Q^2)^{2 i -2}$ ($i > 2$) \cite{data2N,PSTSK10} and even their infinite sums \cite{Teryaev:2013qba,Gabdrakhmanov:2018eop} are considered at low $Q^2$ values. These HT terms increase the accuracy but complicate the corresponding analyses because of the consideration of a set of additional parameters.}}

 \textcolor{black}{It is known that if we use the OPE formalism, we should in principle include the elastic contribution $\Gamma_1^{p-n}(Q^2;{\rm el.})$ coming from $x=1$  in BSR (\ref{BSRdef}). The elastic contribution is expressed  with the electromagnetic form factors \cite{Osip,Merg}, which are available in parametrized form \cite{Merg,Sabbiretal}. The contribution $\Gamma_1^{p-n}(Q^2;{\rm el.})$ is of higher-twist form: at $Q^2 \gtrsim 1 \ {\rm GeV^2}$ we have $\Gamma_1^{p-n}(Q^2;{\rm el.}) \sim 1/(Q^2)^4$. Therefore, by subtracting this contribution from the total BSR, the OPE for the obtained inelastic contributions gets modified only from the dimension $D=8$ term on. We will be using for our theoretical curves the OPE truncated at the $D=2$ term [$\sim 1/Q^2$, cf.~Eq.~(\ref{BSRhigh}) in Sec.~\ref{sec:HT}]. Therefore, there is no compelling reason to include in the fit (to such truncated OPE expression) the experimental points with elastic contribution added. In this context, we also mention that the LT-contribution in the 2$\delta$ and 3$\delta$anQCD does not generate terms with $D \leq 8$, because in these QCD variants the relation (\ref{diff}) is fulfilled with $N=5$, as mentioned earlier. Further, the $Q^2$-dependence of the nonsinglet inelastic BSR in the low-$Q^2$ regime is constrained by the Gerasimov-Drell-Hearn (GDH) sum rule \cite{GDHsr}, as was pointed out in \cite{Ansel,PSTSK10}.  Therefore, we will apply the fit procedures to the experimental points for the pure inelastic contribution, for $Q^2 \gtrsim 1 \ {\rm GeV}^2$ with the truncated OPE, and for low $Q^2$ with GDH-motivated and related ans\"atze \cite{GDHsr,LFH}.}

The twist-2 contribution $E_{\rm {NS}}(Q^2)$ 
in (\ref{BSR}) is known up to N$^3$LO, where the NLO was found 
in \cite{nloBSR}, N$^2$LO in \cite{nnloBSR}, and N$^3$LO contribution
was obtained in \cite{nnnloBSR}. 
These coefficients are presented in Table \ref{tabeNS} 
for various active flavors $N_f$, in $\MSbar$ scheme where we use the notation
\be
E_{\rm NS}(Q^2)=1+e_1^{\rm NS} a_{\rm pt}(Q^2) + 
e_2^{\rm NS} a_{\rm pt}(Q^2)^2 + 
e_3^{\rm NS} a_{\rm pt}(Q^2)^3 + 
e_4^{\rm NS} a_{\rm pt}(Q^2)^4\ .
\label{Ens}
\ee  
\begin{table}
\caption{The nonsinglet coefficients $e_n^{\rm NS}$ (for $n=1,2,3,4$) 
in the expansion (\ref{Ens}) for various flavor numbers $N_f$ 
up to N${^3}$LO order, in $\MSbar$ scheme and for RScl $\mu^2=Q^2$ ($C=0$), for various values of $N_f$.}
\label{tabeNS}
\begin{ruledtabular}
\centering
\begin{tabular}{r|cccc}
%\hline
$N_f$ & $e_1^{\rm NS}$ & $e_2^{\rm NS}$ & $e_3^{\rm NS}$ & $e_4^{\rm NS}$
\\
\hline
3 & -1 & -3.58333 & -20.21527 & -175.7   \\
%\hline
4 & -1 & -3.25000 & -13.85026 & -102.4  \\
%\hline
5 & -1 & -2.91667 & -7.84019 & -41.96 \\
%\hline
6 & -1 & -2.58333 & -2.18506 & 6.2  \\
%\hline
\end{tabular}
\end{ruledtabular}
\end{table}

The perturbative BSR function (\ref{Ens}) depends on a single kinematical variable $Q^2$ .
We can apply the machinery of analytic QCD frameworks, where the perturbation 
series (\ref{Ens}) should be expressed as a nonpower series via the transformation 
$a(Q^2)^n\mapsto\A_n(Q^2)$, and is given by
\be
E_{{\rm NS},j}(Q^2)=1+e_1^{\rm NS} \A^{(j)}(Q^2) + 
e_2^{\rm NS} \A_2^{(j)}(Q^2) + 
e_3^{\rm NS} \A_3^{(j)}(Q^2) + 
e_4^{\rm NS} \A_4^{(j)}(Q^2)\ ,
\label{anEns}
\ee 
where index $j$ indicates in which analytic QCD framework we are 
working ($j=$APT, $2\delta$anQCD, and  $3\delta$anQCD). 
For the numerical evaluation
of $\A_n^{(j)}(Q^2)$'s we use various programs \cite{mathprg,2danQCD,3l3danQCD,4l3danQCD,MathDown}\footnote{
The programs can also be downloaded from the web page \cite{MathDown}.}
written in Mathematica.

In 2$\delta$anQCD and 3$\delta$anQCD, the renormalization scheme ($c_2, c_3, \ldots$) is different from $\MSbar$ scheme (${\overline c}_2, {\overline c}_3, \ldots$), cf.~Appendix \ref{App:A}. As a consequence, the coefficients $e_3^{\rm NS}$ and $e_4^{\rm NS}$ become different in that scheme. If we denote by ${\overline e}_k^{\rm NS}$ the coefficients in $\MSbar$ scheme (i.e., those of Table \ref{tabeNS}), and by ${e}_k^{\rm NS}$ the corresponding coefficients in a different scheme  ($c_2, c_3, \ldots$), the relations between them follow from the scheme independence of the quantity $E_{\rm NS}(Q^2)$
\bes
\label{ebare}
\bea
e_1^{\rm NS} & = & {\overline e}_1^{\rm NS} = -1
\\
e_2^{\rm NS} & = & {\overline e}_2^{\rm NS}
\\
e_3^{\rm NS} & = & {\overline e}_3^{\rm NS}-{\overline e}_1^{\rm NS}(c_2 - {\overline c}_2)
\\
e_4^{\rm NS} & = & {\overline e}_4^{\rm NS}-2 {\overline e}_2^{\rm NS}(c_2 - {\overline c}_2) - {\overline e}_1^{\rm NS} \frac{1}{2}(c_3 - {\overline c}_3).
\eea
\ees
At low $Q^2 \lesssim 1{\rm GeV}^2$ we use $N_f=3$ throughout.

We note that we can also use different renormalization scale (RScl) $\mu^2 \not= Q^2$ in the evaluation of the quantity $E_{{\rm NS},j}(Q^2)$. In such a case, the dependence on the RScl parameter $C \equiv \ln(\mu^2/Q^2)$ enters the coefficients $e_j^{\rm NS} \mapsto e_j^{\rm NS}(C)$ and the couplings $\A_n^{(j)}(Q^2 \exp(C))$. The evaluated expressions $E_{\rm NS,j}(Q^2)$ depend on RScl and on the scheme parameters $c_m$ ($m \geq 2$) because the evaluated series is truncated (at $\sim a_{\rm pt}^4$ or $\A_4^{(j)}$).  

As mentioned earlier, among the HT terms in BSR (\ref{BSR}) we will include in our analysis only the twist-4 term. This term has known evolution \cite{ShuVa,KUY} in pQCD
\be
\mu_{4}^{p-n}(Q^2)=\mu_{4}^{p-n}(Q_{\rm in}^2)\left(\frac{a_{\rm pt}(Q^2)}
{a_{\rm pt}(Q_{\rm in}^2)}\right)^{\gamma_0/8\beta_0}\ ,
\label{HTQ2}
\ee
where the first nonsinglet anomalous dimension coefficient is $\gamma_0=16 C_F/3=64/9$, and we will fix the initial evolution scale at $Q_{\rm in}^2=1 \ {\rm GeV}^2$. When $N_f=3, 4$, we have $\gamma_0/(8\beta_0)=32/81$ and $32/75$, respectively. We note that $\mu_{4}^{p-n}$ contains target mass corrections from twist-2 and twist-3 and the color polarizability.

In general analytic versions of QCD, the powers $a^{\nu}$ (where $\nu$ is not necessarily integer) get transformed to $\A_{\nu}$ (which is in general different from $\A^\nu$), according to the general formalism of Ref.~\cite{GCAK}, cf.~also Eq.~(\ref{AntAn}). We apply it to the twist-4 term (\ref{HTQ2})
\be
\mu_{4,j}^{p-n}(Q^2)=\mu_{4,j}^{p-n}(Q_{\rm in}^2) \frac{\A_{\gamma_0/8\beta_0}^{(j)}(Q^2)}
{\A_{\gamma_0/8\beta_0}^{(j)}(Q_{\rm in}^2)}\ .
\label{HTQ2an}
\ee
In the leading order, which is the case for $\mu_4^{p-n}$, we have $\A_{\nu} = \tA_{\nu}$, cf.~Eq.~(\ref{AntAn}). The relevant programs are available from the web page \cite{MathDown}. 

Only in FAPT, an equivalent, but more direct procedure can be applied for the evaluation of $\A_{\nu}$, namely, it is a dispersion integral containing the discontinuity function ${\rm Im} a_{\rm pt}^{\nu}(-\sigma - i\epsilon)$, Eq.~(\ref{AnuMA}). For a discussion and details about analytization in structure functions of the proton in FAPT QCD (with $\nu$ noninteger), we refer to \cite{SFsA}.

\begin{figure}[htb] 
\begin{minipage}[t]{0.45\textwidth}%
\includegraphics[width=\textwidth]{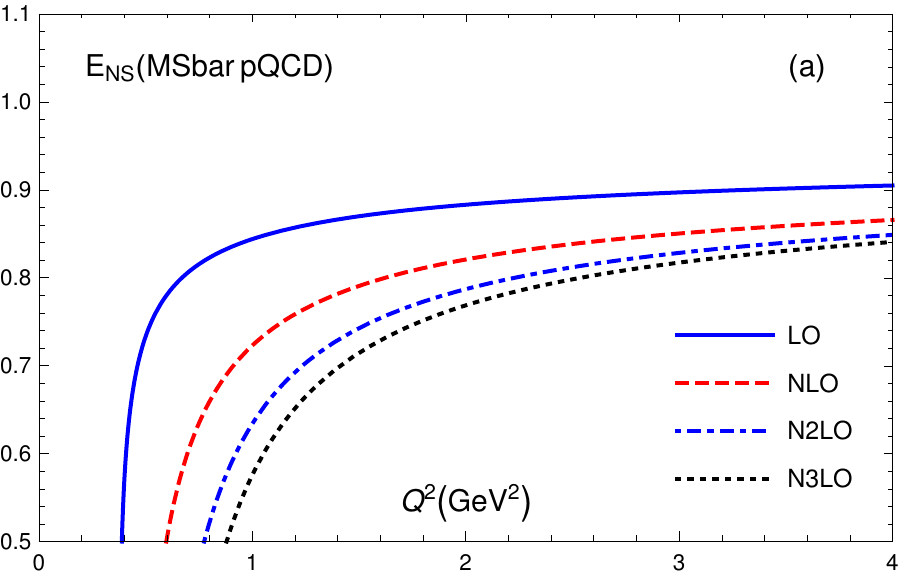}
\end{minipage}
\begin{minipage}[t]{0.45\textwidth}
\includegraphics[width=\textwidth]{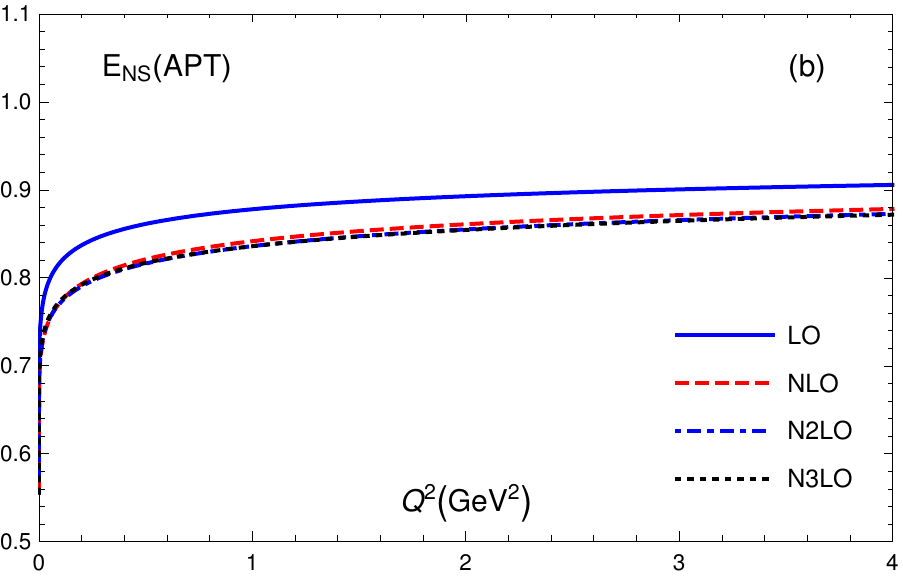}
\end{minipage}
\begin{minipage}[t]{0.45\textwidth}
\includegraphics[width=\textwidth]{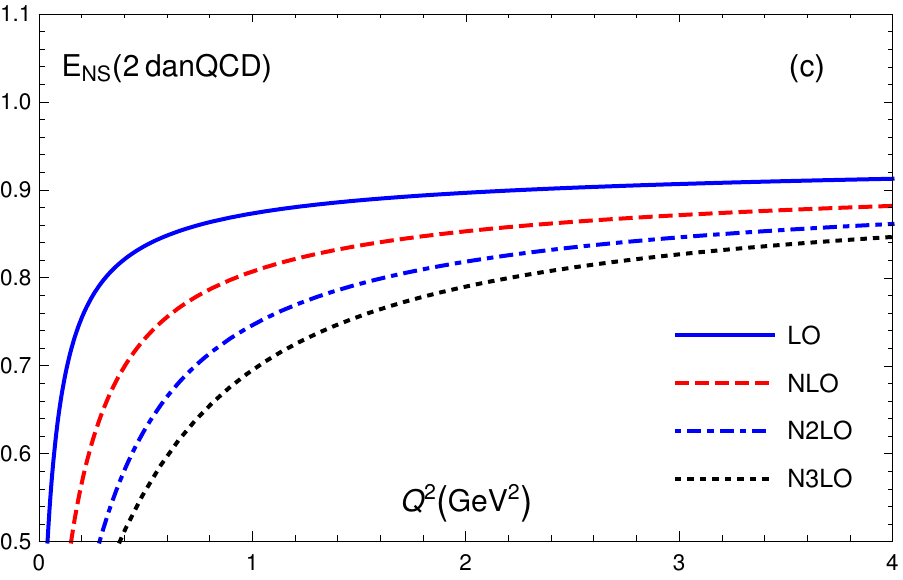}
\end{minipage}
\begin{minipage}[t]{0.45\textwidth}
\includegraphics[width=\textwidth]{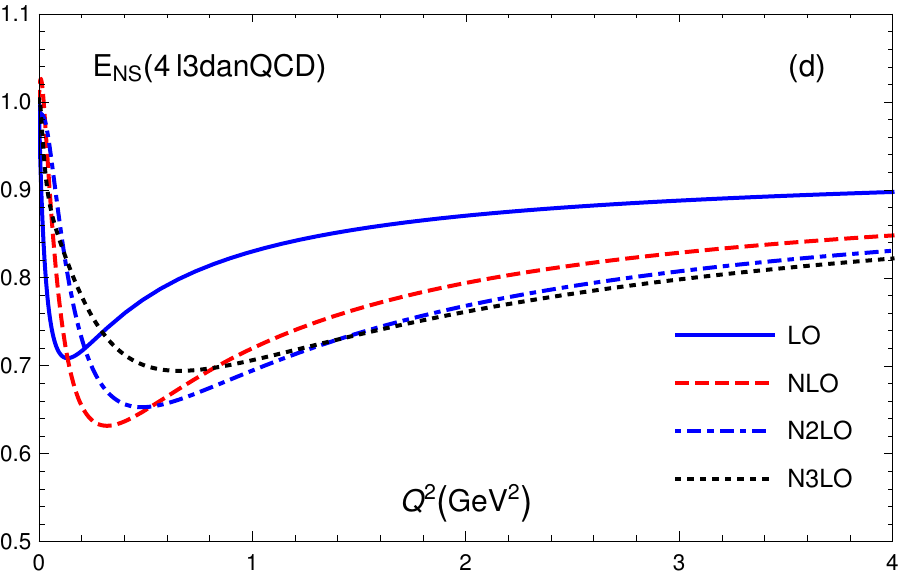}
\end{minipage}
\vspace{-0.2cm}
\caption{(a) The nonsinglet order by order perturbative series (\ref{Ens}) as a function  of $Q^2$ in $\MSbar$ pQCD; and the corresponding series (\ref{anEns}) in the analytic QCD frameworks: (b) APT; (c) 2$\delta$anQCD; and (d) 3$\delta$anQCD. RScl was taken $\mu^2=Q^2$. The underlying pQCD for APT
is in the $\MSbar$ scheme, for 2$\delta$anQCD is in the Lambert scheme with $c_2=-4.9$, and for 3$\delta$anQCD in the four-loop MiniMOM scheme.}
\label{EnsConv}
\end{figure}
It is interesting to analyze the behavior of the perturbation series (\ref{Ens}) and (\ref{anEns}) and their convergence in all frameworks that we are considering. We can see in Figs.~\ref{EnsConv}(a)-(d) the quantity $E_{\rm NS}(Q^2)$ for each order in the perturbation series for: (a) pQCD, (b) APT, (c) 2$\delta$anQCD, and (d) 3$\delta$anQCD (in four-loop MiniMOM). We observe that pQCD has a bad convergence, mainly due to the renormalon ambiguity\footnote{In Ref.~\cite{Pineda05} the first IR renormalon was eliminated in a specific ``renormalon subtracted scheme'' and, as a consequence, the convergence of BSR pQCD series improved.}
 and due to the unphysical behavior of $a_{\rm pt}(Q^2)$ at low $Q^2$ (because of the vicinity to the aforementioned Landau singularities). Unlike pQCD, the analytic frameworks show a clear improvement of both problems that arise in pQCD. APT and  3$\delta$anQCD show faster convergence. When $Q^2$ goes to low values, APT keeps  almost unchanged value down to $Q^2\approx 0.4 \ {\rm GeV}^2$ where it begins to decline rapidly; 2$\delta$anQCD begins to decline at $Q^2 \approx 1.5 \ {\rm GeV}^2$; the lattice-motivated 3$\delta$anQCD first declines somewhat when going down to the region $(0.5, 1)$ GeV, and then increases at low $Q^2 \sim 0.1$ GeV. The fast convergence of the APT nonpower series was noted in \cite{Bakulev},  where for a general noninteger power index $\nu$, very strong hierarchy $\vert\A_{\nu+1}^{(APT)}\vert\ll\vert\A_{\nu}^{(APT)}\vert$ is valid even at very low $Q^2$. In the other anQCD frameworks we do have this hierarchy at all $Q^2$, but it is not as strong as in (F)APT.

\subsection{Low and high-$Q^2$ behavior}
\label{sec:HT}
  
The HT contributions on the right-hand side of Eq.~(\ref{BSR}) are important in the low-energy regime $Q^2\sim 1 \ {\rm GeV}^2$. At very low $Q^2 < 1 \ {\rm GeV}^2$, the HT contribution grows quickly and the OPE series diverges. This is a general problem in OPE. However, we can try to solve this problem if we replace the OPE expression (\ref{BSR}) at low $Q^2 \leq Q_0^2$ ($\approx 0.5 \ {\rm GeV}^2$) with a $\chi$PT-motivated expression \cite{data2B}\footnote{There are alternative methods to deal with this regime, e.g., an extension from the GDH sum rule made via a QCD-improved model \cite{GDHlow1,GDHlow2}, or with a resummation of perturbation series in \cite{Kotikov12}.}
\be
\Gamma_1^{p-n}(Q^2) = \frac{(\kappa_n^2-\kappa_p^2)}{8 M_p^2} Q^2 + A \; (Q^2)^2 + B \; (Q^2)^3
\quad (Q^2 \lesssim 0.5 \ {\rm GeV}^2), 
\label{BSRlow1}
\ee
where $\kappa_X$ is the anomalous moment of the nucleon $X$ ($\kappa_p=1.793$, $\kappa_n=-1.916$), $A$ and $B$ are fit parameters. The first term ($\sim Q^2$) comes from the GDH sum rule \cite{GDHsr}. 

\textcolor{black}{Yet another possibility is to use at low $Q^2 \leq Q_0^2$ ($\lesssim 1 \ {\rm GeV}^2$) in BSR the form of the light-front holographic (LFH) coupling \cite{LFH} in the BSR ($g_1$) scheme \cite{alg1} [$\A(0)_{g_1}=1$]}
\be
\Gamma_1^{p-n}(Q^2) = 
\frac{g_A}{6} \left[ 1 - \A(Q^2)_{\rm LFH} \right] =
\frac{g_A}{6} \left[ 1 - \exp\left( - \frac{Q^2}{4 \kappa^2} \right) \right],
\quad (Q^2 \lesssim 1 \ {\rm GeV}^2). 
\label{BSRlow2}
\ee
\textcolor{black}{Here, $\kappa$ will be the fit parameter. It is expected to be close to the value $\kappa = 0.523 \pm 0.024$ \cite{kappa} characterizing the mass scale of the light-quark hadron spectroscopy.}\footnote{\textcolor{black}{In Ref.~\cite{Brod2}, the coupling $\A(Q^2)_{\rm LFH}$ and its derivative $d \A(Q^2)_{\rm LFH}/d \ln Q^2$ were matched at a scale $Q_0^2$ with the pQCD coupling in $\MSbar$, MOM (in the Landau gauge), V and $g_1$ schemes. These two conditions then fixed the values of $\A(0)_{\rm LFH}$ and the IR/UV transition scale $Q_0^2$. We will not use such a coupling here, but will use $\A(Q^2)_{\rm LFH}$ in $g_1$ scheme only below a scale $Q_0^2 \sim 1 \ {\rm GeV}^2$ according to Eq.~(\ref{BSRlow2}).}}

At higher $Q^2$ we will take OPE (\ref{BSR}) with only one HT term
\be
\Gamma_{1,j}^{p-n} (Q^2) = \frac{g_A}{6} E_{\rm {NS},j}(Q^2) + \frac{\mu_4^{p-n}(Q^2)}{Q^2} \quad (Q^2 \gtrsim 0.5 \ {\rm GeV}^2).
\label{BSRhigh}
\ee
In Sec.~\ref{subs:MSbar}, we will fit to the experimental values only with this OPE, and with LT contribution evaluated only with $\MSbar$ pQCD at different orders, with a view to compare how different orders work.

In Sec.~\ref{subs:combNf3}, we will fit to the experimental values a combination of these two expressions: namely, for $Q^2 > Q_0^2$ the OPE expression (\ref{BSRhigh}) with various couplings of QCD for the LT term; and for $Q^2 < Q_0^2$ the expression (\ref{BSRlow1}) or (\ref{BSRlow2}). At the border value $Q_0^2$ we will impose the condition that the high-$Q^2$ and low-$Q^2$ expressions coincide (match), \textcolor{black}{but the derivatives will not be matched.}\footnote{\textcolor{black}{We note here that there is a regular procedure for imposing the continuity of higher derivatives \cite{GDHlow2}. However, the $\chi$PT-motivated expression here (\ref{BSRlow1}) is expected to start failing already at the scales $Q^2 < Q_0^2$ ($ \approx 0.5 \ {\rm GeV}^2$), and this may limit the possibility to impose the continuity of derivatives.}} This will determine the parameter $B$ in the expression (\ref{BSRlow1}), and hence the free  parameters to fit will be the NP parameters $\mu_4^{p-n}$, $Q_0^2$, and $A$. When using the LFH QCD-motivated expression (\ref{BSRlow2}) at $Q^2 < Q_0^2$, the parameter $\kappa$ will be determined by the mentioned matching at $Q^2=Q_0^2$, and the free parameters to fit will be $\mu_4^{p-n}$ and $Q_0^2$.
In addition, we will vary also the RScl parameter $C \equiv \ln(\mu^2/Q^2)$ in the LT term  $E_{\rm {NS},j}(Q^2)$, this thus representing the additional parameter of the fitting. 

\section{Numerical Results}
\label{sec:num}

\subsection{pQCD with OPE ansatz only}
\label{subs:MSbar}

\begin{figure}[htb] 
\centering\includegraphics[width=122mm]{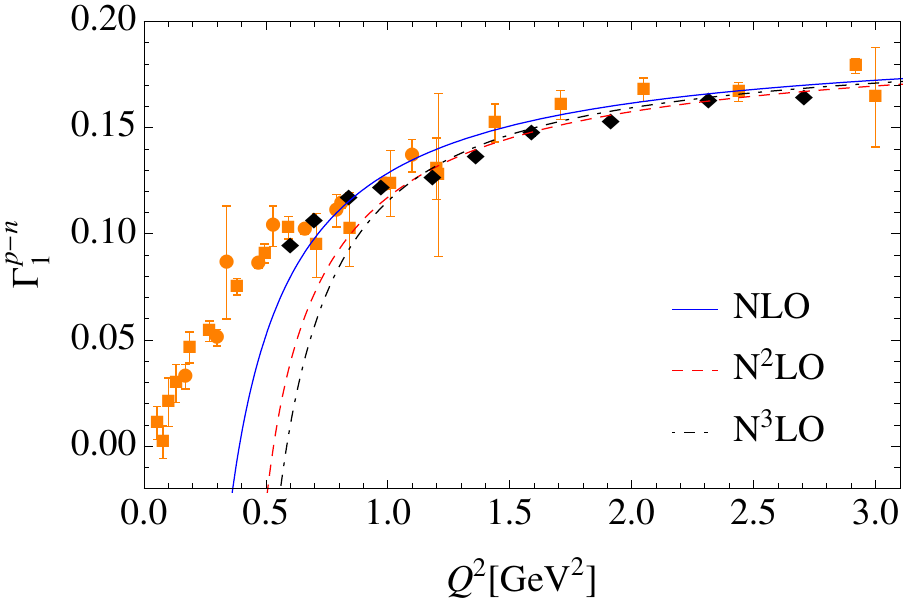} 
\vspace{0.0cm} 
\caption{The $\mu_{4}^{p-n}$ fits of JLAB and SLAC combined data \cite{data2,data2B,data2N,data3} on BSR $\Gamma_1^{p-n}$ as a function of $Q^2$ to various orders of perturbation series (\ref{BSR}): when truncated to $\mathcal{O}$($a^2$) (solid line), to $\mathcal{O}$($a^3$) (dashed line), and to $\mathcal{O}$($a^4$) (dashed-dotted line).  \textcolor{black}{The newer data \cite{data2N} with small statistical errors are in black, and the older data \cite{data2,data2B,data3} are in light grey (orange online).}} \label{figpQCD} 
\end{figure}

First we will test only the OPE approach (\ref{BSRhigh}) with pQCD $\MSbar$
LT term. We take different starting points for the pQCD analysis, since 
the resulting curves in general behave worse when we go to smaller $Q^2$. Therefore, we choose at each order a minimum scale $Q^2=Q^2_{\rm min}$ where $\chi^2$ is minimal (the other fit parameter is $\mu_4^{p-n}$). 
As we are investigating the $Q^2$-dependence of the BSR function, 
we need to fix the $\MSbar$ QCD scale ${\overline \Lambda}$, where we follow the standard extraction, i.e., 
${\overline \Lambda}_{\rm (N_f=3)}$ value comes from a reference value, $a(Q^2=M_Z^2)=0.1185/\pi$ \cite{PDG2014}. We obtain for the pQCD case at LO 
${\overline \Lambda}_{\rm LO}=146$ MeV, at NLO 
${\overline \Lambda}_{\rm NLO}=365$ MeV, at N$^2$LO 
${\overline \Lambda}_{\rm N^2LO}=336$ MeV, and at N$^3$LO 
${\overline \Lambda}_{\rm N^3LO}=344$ MeV. 
The RGE evolution of
$a_{\rm pt}(Q^2)$  from $Q^2=M_Z^2$ ($N_f=5$) 
down to low $Q^2$ where $N_f=3$ is performed at N$^3$LO
with four-loop $\MSbar$ beta function
and with the corresponding three-loop quark threshold conditions \cite{CKS}
(at thresholds $k {\overline m}_q( {\overline m}_q)$ with $k=1$).
At  N$^2$LO, NLO and LO this is performed with the correspondingly 
lower-loop expressions for the beta function and the threshold relations.
\begin{table}
\caption{The HT coefficient $\mu_4^{p-n}(Q^2_{\rm in})$ (where $Q^2_{\rm in}=1 \ {\rm GeV}^2$)  \textcolor{black}{and $Q^2_{\rm min}$ (both in ${\rm GeV}^2$), for $\MSbar$ pQCD with $\mu^2=Q^2$,} extracted from the 
  combined JLAB and SLAC data in various perturbative QCD orders (up to N$^3$LO). \textcolor{black}{Only statistical experimental errors were considered in the fit.}}
\label{tabHTpQ}
\begin{ruledtabular}
\centering
\begin{tabular}{r|cccc}
%\hline
pQCD & LO & NLO & N$^2$LO & N$^3$LO
\\
\hline
 $\mu_{4,pQCD}^{p-n}(1.)$ & -0.059$\pm$0.002 & -0.037$\pm$0.002 & -0.031$\pm$0.002 & -0.008$\pm$0.002 \\
$Q^2_{\rm min}$ & 0.660 & 0.660 & 0.844 & 0.844
%\hline
\end{tabular}
\end{ruledtabular}
\end{table}
In Table \ref{tabHTpQ} we show our obtained values of $\mu_{4,pQCD}^{p-n}$ (where only the statistical 
errors were considered) to various orders in the perturbation expansion (\ref{Ens}). 
As was noted in previous works \cite{PSTSK10,Kataev:1997nc,Narison:2009ag} (see also Ref. \cite{Parente:1994bf}), a duality between HT contribution and the order of the perturbation series appears: when we go to higher order in pQCD, HT contribution decreases in its absolute value. However, this apparent duality property is unstable, since the $\mu_{4,pQCD}^{p-n}$ coefficient is very sensitive to $\Lambda^{(\rm pQCD)}$ parameter at N$^3$LO \cite{PSTSK10}.
The extracted values are consistent with those found at 
LO ($\mu_{4,pQCD}^{p-n}=-0.047\pm0.025 \ {\rm GeV^2}$) in \cite{Ross:1993gb}, 
and at NLO ($\mu_{4,pQCD}^{p-n}=-0.028\pm0.019 \ {\rm GeV^2}$) in \cite{Sidorov:2006vu}. In Fig.~\ref{figpQCD} we see the pQCD fit of BSR function $\Gamma_1^{p-n}$ for NLO, N$^2$LO and N$^3$LO. When we increase the perturbation order, the applicability region of pQCD decreases, covering fewer points of data in the low-$Q^2$ region.

\subsection{Combined analysis, $N_f=3$}
\label{subs:combNf3}

When employing analytic (holomorphic) QCD approaches in the fits,
we recall that the scale parameter ${\overline \Lambda}$ (or $\Lambda_{\rm L.}$)
at $N_f=3$ in the corresponding underlying pQCD coupling
$a_{\rm pt}$ is determined by the condition that the known high-energy
QCD phenomenology be reproduced by such anQCD frameworks.
In 2$\delta$anQCD and 3$\delta$anQCD the $\A(Q^2)$ coupling practically coincides with its underlying pQCD coupling $a_{\rm pt}(Q^2)$ at high $|Q^2|$, i.e., Eq.~(\ref{diff}) is fulfilled with high index value $N=5$. Therefore, in these analytic frameworks we can take such values of $\Lambda_{\rm L.}$ that $a_{\rm pt}(Q^2)$, when converted to the $\MSbar$ scheme and evolved with four-loop RGE\footnote{and with three-loop quark threshold conditions at $Q^2= ( 2 {\overline m}_q({\overline m}_q) )^2$ \cite{CKS}.} to high $Q^2=M_Z^2$, coincides with a typical world average value for $\alpha_s(M_Z^2;\MSbar)$. In this work, we take $\alpha_s(M_Z^2;\MSbar)=0.1185$ \cite{PDG2014}, for our evaluations with 2$\delta$anQCD and 3$\delta$anQCD as well as with $\MSbar$ pQCD. Further, the value of the (leading-twist) $\tau$-lepton decay ratio $r_{\tau}^{(D=0)}$ was set equal to $0.201$, in 3$\delta$anQCD in both the three-loop and the four-loop MiniMOM schemes, as well as in 2$\delta$anQCD. The Mathematica packages for calculation of the couplings $\A_{\nu}(Q^2)$ in all these anQCD versions are freely available \cite{MathDown}.

In (F)APT, the mentioned reasoning about the underlying pQCD coupling $a_{\rm pt}(Q^2)$ does not hold, because at high $|Q^2|$ the (F)APT coupling $\A(Q^2)$ does not coincide with $a_{\rm pt}(Q^2)$ to a high precision. Namely, Eq.~(\ref{diff}) has in this case a low index, $N=1$.
In the global (F)APT, a large part of the QCD phenomenology (low- and
high-energy) is reproduced by the values
${\overline {\Lambda}}_5 \approx 260$ MeV \cite{Sh}
(see also \cite{BMS06,BMS10,Bakulev}),
which is equivalent to $\alpha_s(M_Z^2,\MSbar) \approx 0.122$.
This corresponds to ${\overline {\Lambda}}_3 \approx 400$ MeV
\cite{mathprg} obtained by four-loop RGE approach, 
with three-loop quark thresholds at $Q= 2 {\overline m}_q$ \cite{CKS},
in global APT.
In our analysis we use (F)APT with a fixed value of $N_f=3$ 
[${\rm (F)APT}_{N_f=3}$, ``nonglobal''].
It turns out that in the interval $0 < Q^2 < 3 \ {\rm GeV}^2$  the
value of $\A(Q^2)$ of global APT agrees with $N_f=3$ APT
if in the latter we have ${\overline {\Lambda}}_3 \approx 450$ MeV.
Therefore, we use here the value ${\overline {\Lambda}}_3 \approx 450$ MeV
in ${\rm (F)APT}_{N_f=3}$.

\begin{table}
  \caption{The values of the fit parameters obtained with various approaches to BSR (elastic contribution excluded), with $N_f=3$ \textcolor{black}{and using the $\chi$PT-motivated expression (\ref{BSRlow1}) at $Q^2 \leq Q_0^2$:} HT coefficient $\mu_{4,j}^{p-n}(1. \ {\rm GeV}^2)$ (in ${\rm GeV}^2$), the RScl parameter $C= \ln(\mu^2/Q^2)$, the matching border point $Q_0^2$ (in ${\rm GeV}^2$) and the parameter $A$ (in ${\rm GeV}^{-4}$) of the $\chi$PT-motivated ansatz Eq.~(\ref{BSRlow1}). \textcolor{black}{The penultimate column are the values of $\chi^2$ (where all the 40 experimental points are included). The last column are values of $\chi^2$ at $Q^2 \geq 0.3 \ {\rm GeV}^2$ as explained in the text.}}
\label{TabResNf3}
\begin{ruledtabular}
\centering
\begin{tabular}{r|lllllll}
Approach ($j$) & $\mu_{4,j}^{p-n}(1.)$ & $C$ & $Q_0^2$ & $A$ & $B$ & $\chi^2$ & $\chi^2 (Q^2 \geq 0.3 \ {\rm GeV}^2)$
\\
\hline
$\MSbar$ pQCD     & -0.0344 & 1.801 & 0.646 & 0.658 & -0.840 & 24.44 & 27370 
\\
(F)APT            & -0.0498 & 1.019 & 0.633 & 0.658 & -0.840 & 13.53 & 45.64
\\
2$\delta$anQCD    & -0.0238 & -0.859 & 0.500 & 0.831 & -1.269 & 5.49 & 8.05
\\
(3l)3$\delta$anQCD & -0.0105 & 0.795 & 0.467 & 0.752 & -1.065 & 4.97 & 7.49
\\
(4l)3$\delta$anQCD & -0.0187 & 1.017 & 0.431 & 0.842 & -1.342 & 4.95 & 5.79
\end{tabular}
\end{ruledtabular}
\end{table}
\begin{figure}[htb] %\unitlength=1mm
%\centering\epsfig{file=mb(mu).eps,width=8.cm}
%\begin{minipage}[t]{0.45\textwidth}
\centering\includegraphics[width=140mm,height=90mm]{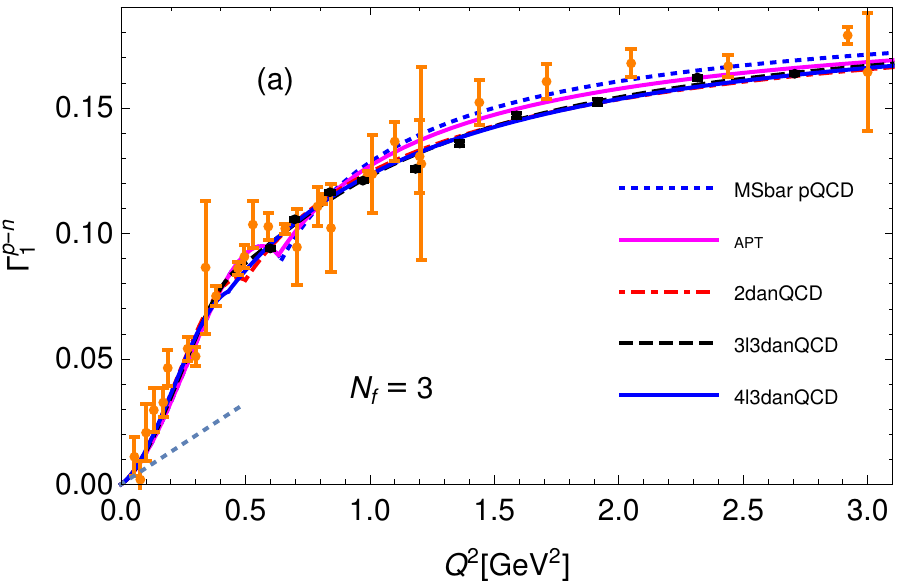}
%\end{minipage}
%\begin{minipage}[t]{0.45\textwidth}
\centering\includegraphics[width=140mm,height=90mm]{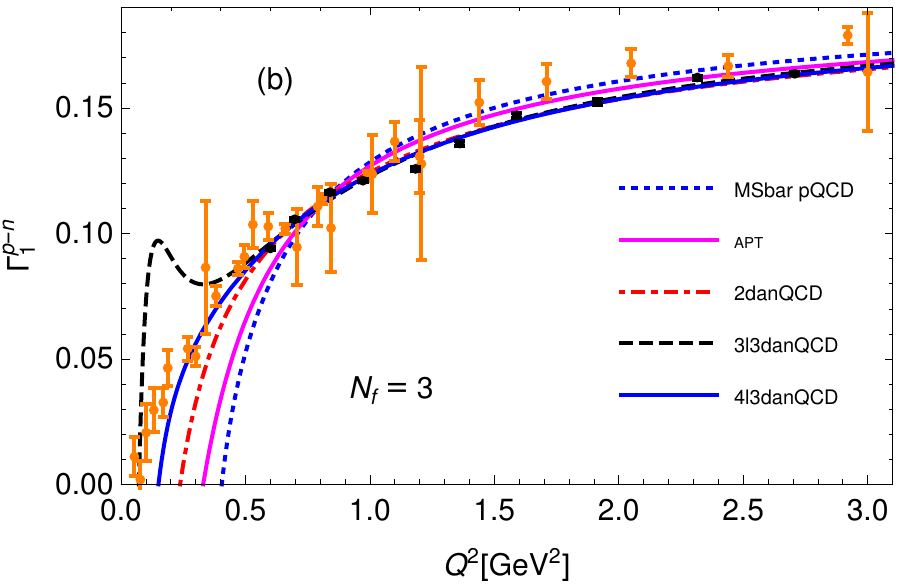}
%\end{minipage}
\vspace{-0.4cm}
\caption{(color online): Fits of JLAB and SLAC data \cite{data2,data2B,data2N,data3} on BSR $\Gamma_1^{p-n}(Q^2)$ (elastic contribution excluded) as a function of $Q^2$:  (a) \textcolor{black}{The combined curves,  at $Q^2 \leq Q_0^2$ the $\chi$PT-motivated expression (\ref{BSRlow1}) and  at $Q^2 \geq Q_0^2$ the OPE curves (\ref{BSRhigh}) with the (four-loop) $\MSbar$ pQCD and holomorphic QCD frameworks}. See the text for details. \textcolor{black}{The GDH sum rule [$~\sim Q^2$ term of Eq.~(\ref{BSRlow1})] is indicated as the dotted straight line.} (b) The OPE curves [with the same values of $\mu_4^{p-n}$ and $C$ as in (a)], but now continued below the point $Q^2=Q_0^2$. \textcolor{black}{The error bars of the experimental points represent statistical errors. The newer data \cite{data2N} with small statistical errors are in black, and the older data \cite{data2,data2B,data3} are in light grey (orange online).}} %
\label{FigFitNf3}
\end{figure}
In Table \ref{TabResNf3} we present, for five different cases of evaluation of the LT contribution $E_{\rm {NS}}(Q^2)$ at $Q^2 \geq Q_0^2$, the obtained values of the fit parameters \textcolor{black}{when using the $\chi$PT-motivated expression (\ref{BSRlow1}) at $Q^2 \leq Q_0^2$:} HT coefficient $\mu_4^{p-n}(Q^2_{\rm in})$; RScl parameter $C \equiv \ln(\mu^2/Q^2)$ of the LT contribution; matching border point $Q_0^2$; parameter $A$ of the $\chi$PT-motivated expression (\ref{BSRlow1}).
%For the RScl $\mu^2$ we imposed the restriction $Q^2/16 < \mu^2 < 16 Q^2$, i.e., $|C| < \ln(16)$ ($=2.773$).\footnote{This restriction was relevant only in the $\MSbar$ pQCD and (F)APT case.}
 The fit for these four parameters was performed with respect to the experimental data; we refer to the previous Section \ref{sec:HT} for more explanation. The values of the parameter $B$ of the $\chi$PT-motivated expression were obtained by the matching condition at $Q^2=Q_0^2$. 
\textcolor{black}{In the penultimate column, the values of $\chi^2$ for the resulting curves are given, where all (i.e., 40) experimental points with $Q^2 \leq 3 \ {\rm GeV}^2$ were included. This is for the combined curves, where the theoretical QCD curves are for $Q_0^2 \leq Q^2 \leq 3 \ {\rm GeV}^2$ and the $\chi$PT-motivated curves are for $Q^2 \leq Q_0^2$. The last column represents the values of  $\chi^2$ from the resulting theoretical QCD curves, but applied in an extended $Q^2$-interval, $0.3 \ {\rm GeV}^2 \leq Q^2 \leq 3 \ {\rm GeV}^2$, where there are 37 experimental points (we recall that the high-$Q^2$ QCD curves were applied in the fitting in the shorter, ``high-$Q^2$'' interval, $Q_0^2 \leq Q^2 \leq 3 \ {\rm GeV}^2$). We point out that the curves at $Q^2 \geq Q_0^2$ were at the four-loop level, using for the LT contribution the ${\rm N}^3{\rm LO}$ expressions (\ref{Ens})-(\ref{anEns}).} 
The obtained values of the parameter $A$ are approximately consistent with the value $A=0.74$ obtained in $\chi$PT calculations in Ref.~\cite{Jietal} but not with the value  $A=2.4$ obtained in Ref.~\cite{Beretal}. The authors of Ref.~\cite{data2B} used pQCD with various HT terms, and obtained for the $\chi$PT-motivated ansatz at low $Q^2 \lesssim 0.5 \ {\rm GeV}^2$ the values $A=0.80$ and $B=-1.13$, similar to ours when $Q_0^2 \approx 0.44$-$0.61 \ {\rm GeV}^2$.

 \textcolor{black}{We point out that the $\chi^2$ values are dominated by the ten newer experimental points \cite{data2N} (in the interval $0.6 \ {\rm GeV}^2 \leq Q^2 < 3 \ {\rm GeV}^2$) because these points have significantly smaller (statistical) errors than the other, older points \cite{data2,data2B,data3}. Only statistical experimental errors are considered in our fits. For the $\delta$anQCD approaches, which in the considered case work better, we imposed the additional condition $Q_0^2 \leq 0.5 \ {\rm GeV}^2$. It turned out that this restriction is automatically fulfilled in 3l3$\delta$anQCD and 4l3$\delta$anQCD, and in 2$\delta$anQCD it increases $\chi^2$ only insignificantly. In $\MSbar$ pQCD and (F)APT, this condition would significantly increase the already large values of $\chi^2$.}
   
The resulting curves are presented in Fig.~\ref{FigFitNf3}(a). We recall that these curves are made up of two curves ``stitched together'' at a matching point $Q_0^2$, namely the OPE curve (\ref{BSRhigh}) for $Q^2 \geq Q_0^2$ and the $\chi$PT-motivated curve (\ref{BSRlow1}) for $Q^2\leq Q_0^2$. In Fig.~\ref{FigFitNf3}(b) we present \textcolor{black}{again the resulting OPE curves (\ref{BSRhigh}) of Fig.~\ref{FigFitNf3}(a), with the same parameters $\mu_4^{p-n}$ and $C$, but now extended below the point $Q^2=Q_0^2$.}

These results indicate that the pQCD $\MSbar$ approach and, to a lesser degree, the (F)APT approach, are not able to avoid a visible kink (slope discontinuity) at $Q^2=Q_0^2$ between the OPE and the $\chi$PT-motivated expression, i.e., to bridge the gap between the high and low-$Q^2$ regimes. On the other hand, 2$\delta$anQCD and 3$\delta$anQCD appear to be able to bridge this gap without a visible kink, cf.~Fig.~\ref{FigFitNf3}(a). Fig.~\ref{FigFitNf3}(b) indicates that 3$\delta$anQCD in the four-loop MiniMOM and, to a lesser degree, 2$\delta$anQCD, describe the BSR experimental data reasonably well even in the low-$Q^2$ region $Q^2<Q_0^2$ where ($\MSbar$) pQCD approach fails entirely.
\begin{table}
  \caption{\textcolor{black}{As in Table \ref{TabResNf3}, but with the low-$Q^2$ expression (\ref{BSRlow2}).}}
\label{TabResNf3B}
\begin{ruledtabular}
\centering
\begin{tabular}{r|llllll}
Approach ($j$) & $\mu_{4,j}^{p-n}(1.)$ & $C$ & $Q_0^2$ & $\kappa$ & $\chi^2$ & $\chi^2 (Q^2 \geq 0.3 \ {\rm GeV}^2)$
\\
\hline
$\MSbar$ pQCD     & -0.0345 & 1.701 & 0.904 & 0.520 & 20.76 &  127600
\\
(F)APT            & -0.0497 & 0.938 & 0.810 & 0.516 & 14.10 & 44.29 
\\
2$\delta$anQCD    & -0.0238 & -0.869 & 0.584 & 0.504 & 5.67 & 8.02
\\
(3l)3$\delta$anQCD & -0.0105 & 0.645 & 0.705 & 0.503 & 4.35 & 11.92
\\
(4l)3$\delta$anQCD & -0.0187 & 1.016 & 0.300 & 0.494 & 4.90 & 5.79
\end{tabular}
\end{ruledtabular}
\end{table}
\begin{figure}[htb] %\unitlength=1mm
%\centering\epsfig{file=mb(mu).eps,width=8.cm}
%\begin{minipage}[t]{0.45\textwidth}
\centering\includegraphics[width=140mm,height=90mm]{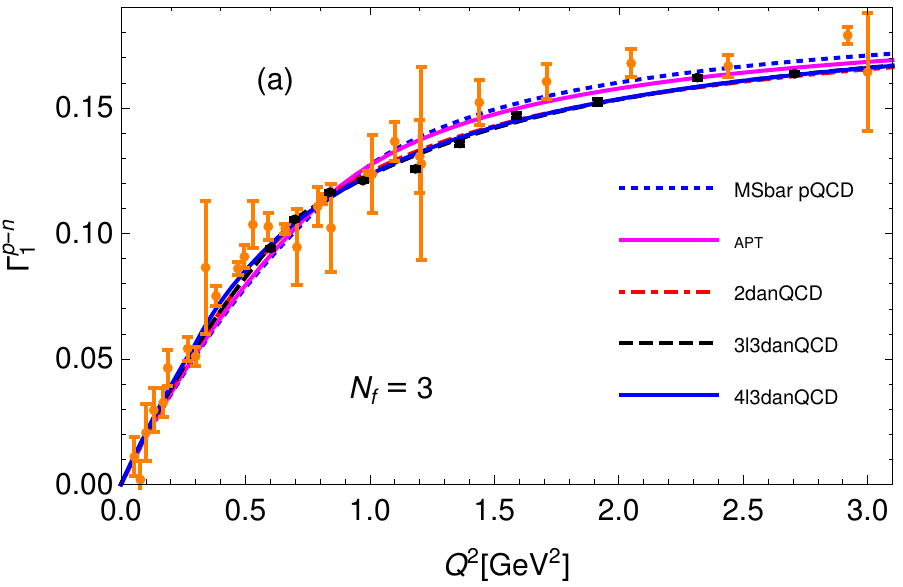}
%\end{minipage}
%\begin{minipage}[t]{0.45\textwidth}
\centering\includegraphics[width=140mm,height=90mm]{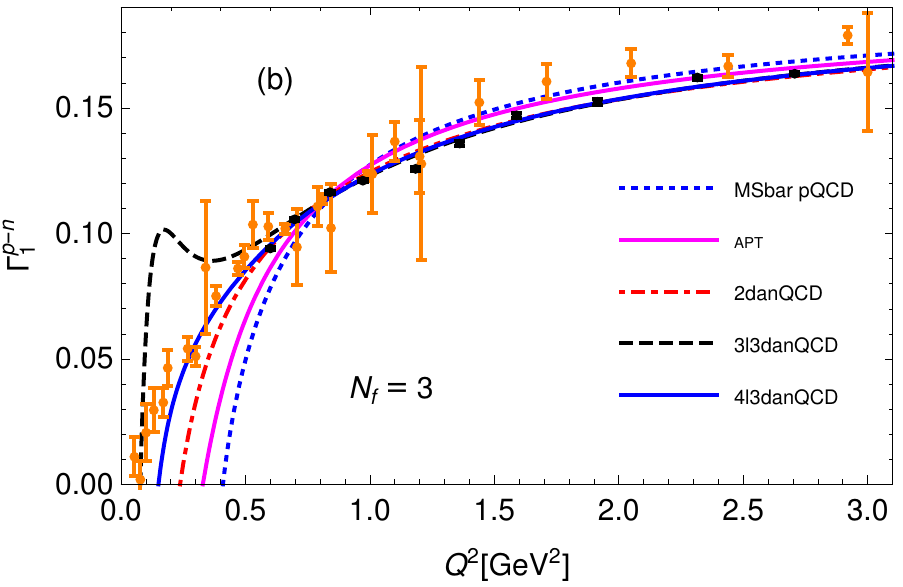}
%\end{minipage}
\vspace{-0.4cm}
\caption{\textcolor{black}{(color online): As Figs.~(\ref{FigFitNf3}), but now the low-$Q^2$ expression has the form of Eq.~(\ref{BSRlow2}) instead of Eq.~(\ref{BSRlow1}).}}
\label{FigFitNf3B}
\end{figure}

\textcolor{black}{We repeat the same type of analysis, but this time with the LFH QCD-motivated ansatz (\ref{BSRlow2}) at $Q^2 \leq Q_0^2$. This time we have only three fit parameters, namely $\mu_4^{p-n}$, $C$, and $Q_0^2$. The $\kappa$ parameter of the expression (\ref{BSRlow2}) is then fixed by the matching condition, i.e., by stitching together the high-$Q^2$ QCD curves and the expression (\ref{BSRlow2}) at $Q^2=Q_0^2$. The results of this analysis are given in Table \ref{TabResNf3B} and in Figs.~\ref{FigFitNf3B}. We note that these results are similar to those of Table  \ref{TabResNf3} and  Figs.~\ref{FigFitNf3}, but are even somewhat better because now the discontinuity in the slope at $Q^2=Q_0^2$ is practically invisible in all cases, i.e., including APT and $\MSbar$ pQCD. Furthermore, the extracted values of the $\kappa$ parameter, $\kappa \approx 0.5 \ {\rm GeV}^2$, are consistent with the value  $\kappa = 0.523 \pm 0.024$ \cite{kappa} obtained from the light-quark hadron spectroscopy in the light-front holographic (LFH) QCD approach with an effective supersymmetric QCD light-front Hamiltonian.}

\textcolor{black}{In both cases, i.e., when using in the low-$Q^2$ regime the $\chi$PT-motivated ansatz (\ref{BSRlow1}) or the LFH QCD-motivated ansatz (\ref{BSRlow2}), we see that the curves with (4-loop) 3$\delta$anQCD approach for $Q^2 \geq Q_0^2$ have the best quality, because $\chi^2$ and $\chi^2(Q^2 \geq 0.3 {\rm GeV}^2)$ are smaller and $Q_0^2$ values are low.}

If we took at $Q^2> {\overline m}_c^2$ ($\approx 1.6 \ {\rm GeV}^2$) for the number of active quarks $N_f=4$, the values of the couplings there would change somewhat and a new, singlet, contribution at $\sim a^4$ would appear.
This will be looked into in the following Section \ref{subs:combNf34}.
 
\subsection{Combined analysis, $N_f=3$-$4$.}
\label{subs:combNf34}

Since the curves in the previous Sections cover the values up to $Q^2=3 \ {\rm GeV}^2$, we have an option to replace at $Q^2 > {\overline m}_c({\overline m}_c)^2$ ($=1.27^2 \ {\rm GeV}^2$) the number of active quarks $N_f=3$ by $N_f=4$. In that case there is an additional singlet contribution at $\sim a_{\rm pt}^4$ term, and the perturbation series (\ref{Ens}) gets replaced by
\bea
E_{\rm pt}(Q^2;N_f)
%E_{\rm NS}(Q^2)
&=& 1+e_1^{\rm NS}(N_f) a_{\rm pt}(Q^2;N_f) + 
e_2^{\rm NS}(N_f) a_{\rm pt}(Q^2;N_f)^2 + 
e_3^{\rm NS}(N_f) a_{\rm pt}(Q^2;N_f)^3
\nonumber \\
&&+ 
\left(e_4^{\rm NS}(N_f) + 3 \; {\rm Tr} [{\cal E}(N_f)]  \; e_4^{\rm SI}(N_f) \right)   a_{\rm pt}(Q^2;N_f)^4\ ,
\label{EnsNf4}
\eea  
where, following Ref.~\cite{Baikov:2015tea} ${\cal E}(N_f)=diag(e_f)$ is the quark charge $e_f$ matrix: ${\rm Tr}{\cal E}(3)=0$, ${\rm Tr}{\cal E}(4)=2/3$.
The singlet coefficient has the following form \cite{Baikov:2015tea}:
\be
e_4^{\rm SI}(N_f) = \frac{\beta_0}{9} d^{abc} d^{abc} 
\label{si}
\ee
where $\beta_0=(11 - 2 N_f/3)/4$, and $d^{abc} d^{abc}=40/3$ (see also \cite{Larin:2013yba}).\footnote{The structure $\beta_0 d^{abc} d^{abc}$ of the result for $e_4^{\rm SI}(N_f)$ was predicted earlier in Ref.~\cite{Larin:2013yba} \textcolor{black}{, where a modification of the generalized Crewther relations of Ref.\cite{GenCrewRel1} was used; the generalized Crewther relations were studied in Refs.~\cite{GenCrewRel2,nnnloBSR}.}} We note that the (underlying) pQCD couplings $a_{\rm pt}(Q^2;3)$ and $a_{\rm pt}(Q^2;4)$ are related by the (3-loop) threshold relation \cite{CKS} at the threshold energy $Q^2 = (k {\overline m}_c)^2$, where $k \sim 1$, and we denote ${\overline m}_c \equiv {\overline m}_c({\overline m}_c)$ ($=1.27$ GeV) the $\MSbar$ mass of $c$ quark. We used $k=2$ in all the cases, for $\MSbar$ pQCD and for the underlying pQCD couplings of the analytic QCD frameworks. We will now introduce in the LT BSR $E(Q^2)$ the $N_f$-dependence in the following form:
\bea
E(Q^2) & = & {\Bigg \{}
\begin{array}{c}
  E(Q^2;N_f=3) \quad (Q^2 < {\overline m}_c^2) \\
  E(Q^2;N_f=4) \quad (Q^2 > {\overline m}_c^2)
\end{array}
.
\label{Nf34}
\eea
We will take this prescription to be independent of the RScl parameter $C = \ln(\mu^2/Q^2)$ used in the couplings and coefficients of the expansion. We point out that, with such an approach, we expect the BSR $\Gamma^{p-n}(Q^2)$ to show a discontinuity at $Q^2={\overline m}_c^2$ ($=1.613 \ {\rm GeV}^2$), principally because the ${\rm N}^3{\rm LO}$ coefficient $e_4(N_f)$ has a discontinuity when $N_f=3 \mapsto 4$, and because the couplings $a_{\rm pt}(Q^2 e^C; N_f)$ and thus also $\A_n(Q^2 e^C;N_f)$ have discontinuities for such $Q^2$ [$\A_n({\overline m}_c^2 e^C;3) \not= \A_n({\overline m}_c^2 e^C;4)$]. We recall that in analytic frameworks we replace in the perturbation series the powers $a_{\rm pt}(Q^2 e^C; N_f)^n$ by $\A_n(Q^2 e^C; N_f)$, cf.~also Eqs.~(\ref{Ens}) and (\ref{anEns}).

The analytic frameworks 2$\delta$anQCD and 3$\delta$anQCD at $N_f=4$ are constructed in such a way as to maintain the pQCD condition $\A(\mu^2) - a_{\rm pt}(\mu^2) \sim (\Lambda^2/\mu^2)^5$ [i.e., Eq.~(\ref{diff}) with $N=5$] not only in the $N_f=3$ region, but also at $\mu^2 > (2 {\overline m}_c)^2$, i.e., in the $N_f=4$ region. Further, in the $N_f=4$ coupling in the (lattice-motivated) 3$\delta$anQCD we formally keep the $Q^2=0$ condition $\A(Q^2=0;N_f=4)=0$, although this condition is optional for $N_f=4$. In practice, we kept in 2$\delta$anQCD($N_f=4$) the same value of $c_2=-4.9$ and the same value of $s_0=25.61$ as in the $N_f=3$ case (cf.~Table \ref{tab2dan}). In 3l3$\delta$anQCD($N_f=4$) (three-loop MiniMOM scheme) we used the value $c_2(N_f=4)$ of the MiniMOM scheme, and kept the same values of the parameters $s_0\equiv M_0^2/\Lambda_{\rm L.}^2$ ($=3.00$) and $f_1 \equiv {\cal F}_1/\Lambda^2_{\rm L.}$ ($=0.04537$) as in 3l3$\delta$anQCD($N_f=3$). In 4l3$\delta$anQCD($N_f=4$) (four-loop MiniMOM scheme) we used the values $c_2(N_f=4)$ and $c_3(N_f=4)$ of the MiniMOM scheme, and kept the same values of the parameters $s_0$ ($=652$) and $s_1 \equiv M_1^2/\Lambda_{\rm L.}^2$ ($=3.97$) as in 4l3$\delta$anQCD($N_f=3$).

\begin{figure}[htb] %\unitlength=1mm
%\centering\epsfig{file=mb(mu).eps,width=8.cm}
%\begin{minipage}[t]{0.45\textwidth}
\centering\includegraphics[width=140mm,height=90mm]{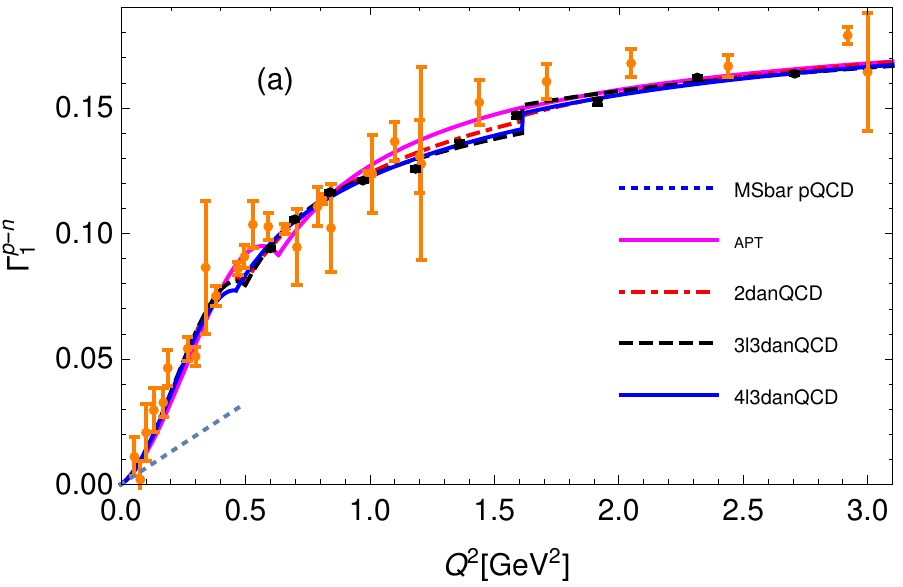}
%\end{minipage}
%\begin{minipage}[t]{0.45\textwidth}
\centering\includegraphics[width=140mm,height=90mm]{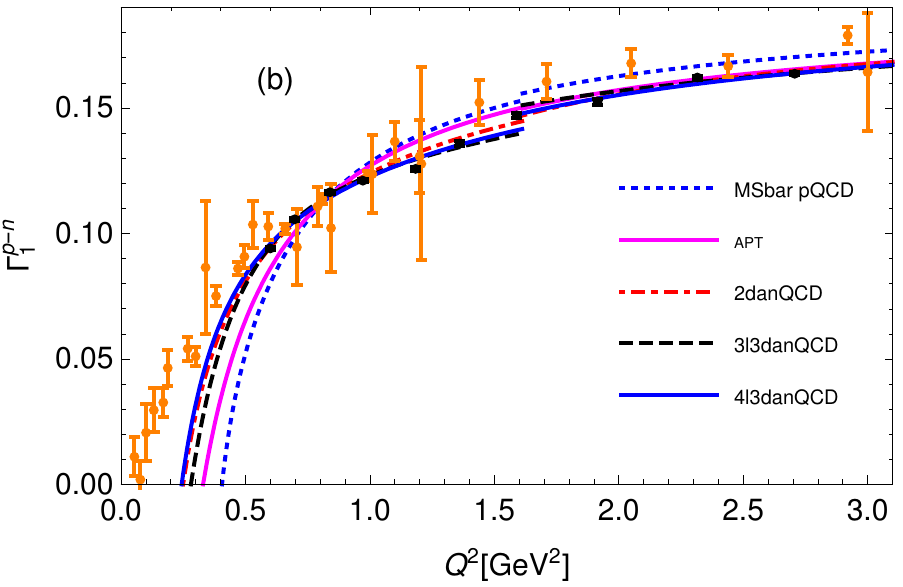}
%\end{minipage}
\vspace{-0.4cm}
\caption{As Figs.~\ref{FigFitNf3}, but now with $N_f=4$ at $Q^2 > {\overline m}_c^2$.}
\label{FigFitNf34}
\end{figure}

\begin{table}
\caption{As Table \ref{TabResNf3}, but now the fit is performed at $Q^2>{\overline m}_c^2$ with $N_f=4$ theoretical curves.}
\label{TabResNf34}
\begin{ruledtabular}
\centering
\begin{tabular}{r|lllllll}
Approach ($j$) & $\mu_{4,j}^{p-n}(1.)$ & $C$ & $Q_0^2$ & $A$ & $B$ & $\chi^2$  & $\chi^2 (Q^2 \geq 0.3 \ {\rm GeV}^2)$
\\
\hline
$\MSbar$ pQCD      & -0.0362 & 1.968 & 0.647 & 0.658 & -0.840 & 26.85 & 30080
\\
(F)APT             & -0.0498 & 1.017 & 0.633 & 0.658 & -0.840 & 13.36& 45.06
\\
2$\delta$anQCD     & -0.0257 & -1.259 & 0.500 & 0.838 & -1.289 & 4.46 & 7.46
\\
(3l)3$\delta$anQCD & -0.0307 & -0.742 & 0.500 & 0.858 & -1.340 & 4.05 & 10.09
\\
(4l)3$\delta$anQCD & -0.0294 & -0.261 & 0.462 & 0.842 & -1.342 & 4.08 & 5.90
\end{tabular}
\end{ruledtabular}
\end{table}
When repeating the analysis of the previous Section, \textcolor{black}{with the $\chi$PT-motivated expression (\ref{BSRlow1}) at $Q^2 \leq Q_0^2$,} but now with the condition (\ref{Nf34}), we obtain the results presented in Figs.~\ref{FigFitNf34}(a), (b) and in Table \ref{TabResNf34}, in close analogy with the $N_f=3$ results Figs.~\ref{FigFitNf3}(a), (b) and Table \ref{TabResNf3} of the previous Section.
Comparing these Figures, we can see that the introduction of the $N_f=4$ effects at $Q^2 >{\overline m}_c^2$ ($\approx 1.61 \ {\rm GeV}^2$) raises somewhat the curves there and makes the agreement with the experimental points there somewhat better in some of the cases. While the resulting values of the fit parameters are similar to those of the $N_f=3$ case, there are some differences in the values of $\chi^2$.

\begin{figure}[htb] %\unitlength=1mm
%\centering\epsfig{file=mb(mu).eps,width=8.cm}
%\begin{minipage}[t]{0.45\textwidth}
\centering\includegraphics[width=140mm,height=90mm]{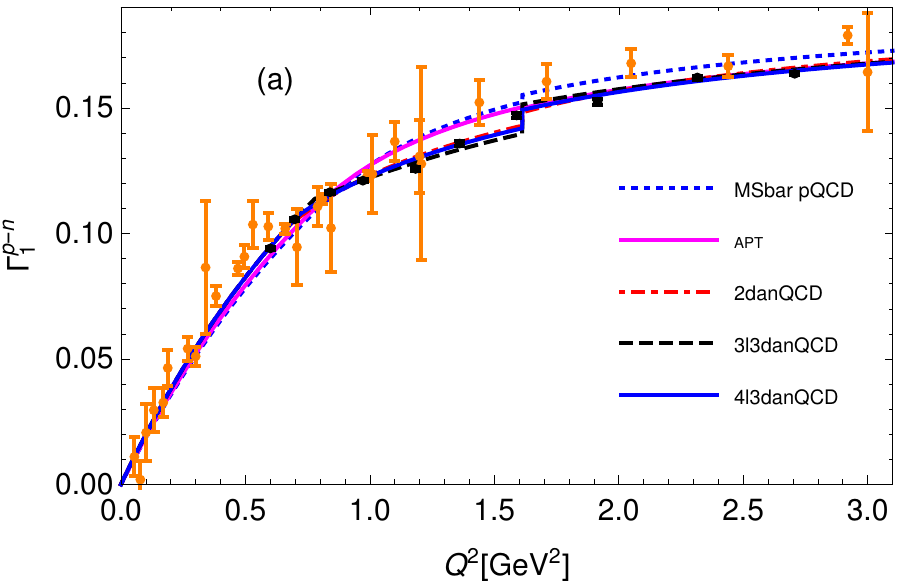}
%\end{minipage}
%\begin{minipage}[t]{0.45\textwidth}
\centering\includegraphics[width=140mm,height=90mm]{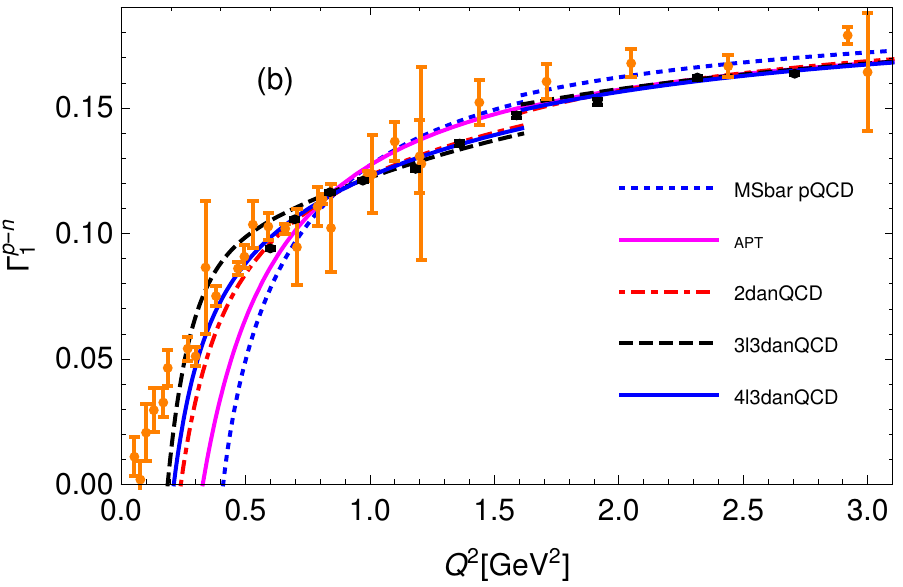}
%\end{minipage}
\vspace{-0.4cm}
\caption{\textcolor{black}{(color online): As Figs.~\ref{FigFitNf34},  but now the low-$Q^2$ expression has the form of Eq.~(\ref{BSRlow2}) instead of Eq.~(\ref{BSRlow1}).}}
\label{FigFitNf34B}
\end{figure}

\textcolor{black}{As in Sec.~\ref{subs:combNf3}, we repeat the same type of analysis, but this time with the LFH coupling ansatz (\ref{BSRlow2}) at $Q^2 \leq Q_0^2$. The results of this analysis are given in Table \ref{TabResNf34B} and in Figs.~\ref{FigFitNf34B}. Again, we note that these results are similar to the results of Tables  \ref{TabResNf34} and  Figs.~\ref{FigFitNf34}, but are even somewhat better because now the discontinuity in the slope at $Q^2=Q_0^2$ is practically invisible in all cases.}
\begin{table}
\caption{\textcolor{black}{As in Table \ref{TabResNf34}, but with the low-$Q^2$ expression (\ref{BSRlow2}).}}
\label{TabResNf34B}
\begin{ruledtabular}
\centering
\begin{tabular}{r|llllll}
Approach ($j$) & $\mu_{4,j}^{p-n}(1.)$ & $C$ & $Q_0^2$ & $\kappa$ & $\chi^2$ & $\chi^2 (Q^2 \geq 0.3 \ {\rm GeV}^2)$
\\
\hline
$\MSbar$ pQCD     & -0.0376 & 2.004 & 0.906 & 0.521 & 23.05 &  32580
\\
(F)APT            & -0.0496 & 0.925 & 0.809 & 0.516 & 13.91 & 43.74 
\\
2$\delta$anQCD    & -0.0249 & -2.244 & 0.679 & 0.504 & 3.90 & 5.86
\\
(3l)3$\delta$anQCD & -0.0214 & -0.217 & 0.779 & 0.503 & 3.85 & 23.12
\\
(4l)3$\delta$anQCD & -0.0260 & 0.113 & 0.717 & 0.503 & 3.55  & 5.99
\end{tabular}
\end{ruledtabular}
\end{table}

\subsection{Comments on our results}
\label{subs:comm}

In all our curves except (F)APT, the reference value $\alpha_s(M_Z^2; \MSbar)=0.1185$ was used. One may ask what happens if this reference value (or the corresponding $\Lambda$ values) are changed. It turns out that the changes in $\chi^2$ are not very significant. For example, if using (in the $N_f=3$ approach) 3l3$\delta$anQCD with the reference value $\alpha_s(M_Z^2; \MSbar)=0.1181$ (the central value of the world average for 2016, \cite{PDG2016}), \textcolor{black}{and the $\chi$PT-motivated expression for $Q^2 \leq Q_0^2$,} then the quality of fit does not change significantly, \textcolor{black}{we obtain $\chi^2=4.92$ instead of $4.97$ (cf.~4th line of Table \ref{TabResNf3}), and $C\equiv \ln(\mu^2/Q^2) =0.841$ instead of $C=0.795$. The other parameters also do not change very significantly: $\mu_4^{p-n}(1.{\rm GeV}^2)=-0.0188 \ {\rm GeV}^2$; $Q_0^2=0.418 \ {\rm GeV}^2$; $A=0.842$ ($\Rightarrow B=-1.342$).}

%The fitting procedure gave in in the case of 2$\delta$anQCD also another solution for the parameters, with a slightly lower value of $\chi^2$ and a large value of $A \approx 1.6$. It turned out that such a solution has at the matching point $Q^2=Q_0^2$ strong discontinuity in the slope, in contrast to the solution presented here. Therefore, such a solution with large $A$ was discarded.

All presented anQCD curves show good results, with the exception of (F)APT \textcolor{black}{which has relatively high values of $\chi^2$ and $\chi^2(Q^2 \geq 0.3{\rm GeV}^2)$.} A possible problem with (F)APT also appears in Figs.~\ref{FigFitNf3}(a) and \ref{FigFitNf34}(a) where we see that there is a visible kink (discontinuity in the first derivative) for the combined curve (F)APT and $\chi$PT-motivated curve (\ref{BSRlow1}), at that $Q^2=Q_0^2$. The ($\MSbar$) pQCD curve also has such a problem, in an even stronger form. One may ask whether the problem of a strong kink in (F)APT can be eliminated or reduced, by reducing the scale $\Lambda_{N_f}$. However, when we reduce $\Lambda_3$ from $0.450$ GeV to $0.400$ GeV, no improvement is obtained (when the $\chi$PT-motivated curve is used for $Q^2 \leq Q_0^2$), the strong kink remains, and (in the $N_f=3$ case) \textcolor{black}{we obtain $Q_0^2=0.635 \ {\rm GeV}^2$ and $\chi^2=15.17$, instead of $Q_0^2=0.633 \ {\rm GeV}^2$ and $\chi^2=13.53$ (cf.~2nd line of Table \ref{TabResNf3})}; also in the combined case $N_f=3$-$4$ the changes due to $\Lambda_3=0.45 \ {\rm GeV} \mapsto 0.40 \ {\rm GeV}$ in (F)APT are insignificant. \textcolor{black}{When we apply the LFH QCD-motivated expression (\ref{BSRlow2}) in the low-$Q^2$ regime, the problem with kinks practically disappears. In such a case, in general, the values of $Q_0^2$ are elevated (especially in the APT and $\MSbar$ pQCD cases, to about $0.8$-$0.9 \ {\rm GeV}^2$). This is so because the LFH QCD-motivated expression (\ref{BSRlow2}) fits better the low-$Q^2$ regime of BSR experimental data than the $\chi$PT-motivated expression (\ref{BSRlow1}).}

We wish to comment also on one particular feature. The discontinuity of the (F)APT curve at  $Q^2 = {\overline m}_c^2$ in the case of $N_f=3$-$4$, Figs.~\ref{FigFitNf34} and \ref{FigFitNf34B}, is practically invisible \textcolor{black}{(in the case of Figs.~\ref{FigFitNf34} it is about one sixth of the discontinuity of the $\MSbar$ pQCD curve)}. Further, \textcolor{black}{comparison of Table \ref{TabResNf34} with \ref{TabResNf3}, and  Table \ref{TabResNf34B} with \ref{TabResNf3B}}, shows that the introduction of $N_f=4$ effects changes the parameters in the (F)APT case insignificantly. This is due to a conjunction of two effects in (F)APT: the discontinuities of the LT and HT contributions at $Q^2= {\overline m}_c^2$ are already small, and these two discontinuities have different signs and result in a relatively strong cancellation.

The authors of Ref.~\cite{BFF} calculated, among other things, the corrections to BSR from heavy quarks (primarily $c$ quark) at $\sim a_{\rm pt}^2$ level. Their results show that (at $\sim a_{\rm pt}^2$), if considering expansion of $E_{\rm pt}(Q^2)$ in powers of $a_{\rm pt}(Q^2;N_f=4)$, the effective number of flavors in the NLO coefficient $e_2^{\rm NS}(N_f^{\rm eff})$ is approximately $3.13, 3.36, 3.73$ for $Q^2=5, 10, 50 \ {\rm GeV}^2$, respectively. Further, if considering expansion in powers of $a_{\rm pt}(Q^2;N_f=3)$, the effective number of flavors in $e_2^{\rm NS}(N_f^{\rm eff})$ is $N_f^{\rm eff} \approx 3$ for $Q^2 < 3 \ {\rm GeV}^2$. We refer for some details to Appendix \ref{App:B}. These results indicate that the $N_f=4$ effects in BSR set in at considerably higher $Q^2$ than $Q^2={\overline m}_c^2$ ($\approx 1.6 \ {\rm GeV}^2$) used here in Sec.~\ref{subs:combNf34}, and that the $N_f=3$ approach (Sec.~\ref{subs:combNf3})  should be a good approximation in the range $Q^2 < 3 \ {\rm GeV}^2$ considered here. 

\textcolor{black}{We recall that one of the presented anQCD approaches, namely $3\delta$anQCD \cite{3l3danQCD,4l3danQCD}, has zero value of the coupling $\A(Q^2)$ at $Q^2=0$, and two of the presented anQCD approaches, APT \cite{ShS,MS96,ShS98,Sh} and 2$\delta$anQCD \cite{2danQCD}, have (finite) nonzero values of the coupling $\A(Q^2)$ at $Q^2=0$: $\A(0)_{\rm APT}=4/9=0.444$ ($=1/\beta_0$ with $N_f=3$); $\A(0)_{\rm 2\delta} \approx 0.66$. This is to be compared with the value of the effective coupling of BSR ($g_1$) scheme \cite{alg1} which is by definition $\A(0)_{g_1}=1$; the latter normalization was used for the IR-safe light-front holographic coupling \cite{LFH} $\A(Q^2)_{\rm LFH} \propto \exp(-Q^2/(4 \kappa^2))$ where $\kappa \approx 0.5$ GeV is obtained from low-energy QCD phenomenology, \textcolor{black}{cf.~also Eq.~(\ref{BSRlow2})}. We refer to \cite{Brodrev} for a review of approaches with various kinds of QCD couplings. Couplings with the condition $\A(0)=0$ other than those of Refs.~\cite{3l3danQCD,4l3danQCD} had been constructed in Refs.~\cite{ArbZaits,Boucaud,mes2}. A construction and use of a holomorphic coupling infinite at the origin is given in \cite{Nest1}.}

%\subsection{Additional comments}
%\label{subs:addcomm}

\textcolor{black}{There are several unsettled theoretical questions involved in the applied (theoretical) frameworks, especially at $Q^2 \geq Q_0^2$. Further, there are possibilities to apply other frameworks and approaches to our analysis. One such possibility would be to apply to BSR at $Q^2 \geq Q_0^2$ the Principle of Maximal Conformality (PMC) \cite{PMC} or a related sequential BLM method \cite{secBLM}. Both methods are extensions of the Brodsky-Lepage-Mackenzie (BLM) scale-setting procedure \cite{BLM} beyond NLO. These approaches fix the scales at each order in such a way that the contributions from the $\beta$-dependent parts of the perturbation coefficients are absorbed into the (powers of the) QCD coupling. Such methods have several attractive features for us: (a) they give results independent of the initial chosen renormalization scale; (b) the results do not have the renormalon-like ($\sim n!$) growth of the perturbation coefficients $e_n$; (c) some of the scales in these approaches may become quite low and thus require the use of IR-safe coupling (such as, for example, the holomorphic couplings applied here), cf.~also Ref.~\cite{BdecPMC,BSRPMC}. In the present work, the renormalization scales in Secs.~\ref{subs:combNf3} and \ref{subs:combNf34} were fixed by numerical fitting (minimization of $\chi^2$) to BSR data, not by theoretical arguments. For all these reasons, it would be interesting to perform in the future an analysis of the BSR sum rules by applying (at $Q^2 \geq Q_0^2$) the scale-setting procedures of PMC and sequential BLM approaches \cite{PMC,secBLM} with pQCD and various holomorphic couplings, and to compare the obtained results to those in Ref.~\cite{BSRPMC}.}

\section{Conclusions and outlook}
\label{sec:concl}

In this work we investigated the behavior of the Bjorken polarized sum rule (BSR) $\Gamma_1^{p-n}(Q^2)$, with the elastic contribution excluded, as a function of squared momentum transfer $Q^2$, at low and moderate $Q^2$ in various QCD approaches, comparing it with the available experimental results. The theoretical expressions used were, for $Q^2 \geq Q_0^2$ ($\approx 0.3$-$0.9 \ {\rm GeV}^2$), the leading-twist (LT) contribution to the presently available order $a_{\rm pt}^4$ plus one higher-twist (HT) term $\mu_4^{p-n}/Q^2$, Eqs.~(\ref{Ens})-(\ref{anEns}) and (\ref{BSRhigh}). At low $Q^2 \leq Q_0^2$, we used either the $\chi$PT-motivated expression (\ref{BSRlow1}) \textcolor{black}{or the LFH QCD-motivated expression (\ref{BSRlow2}).} The fit parameters were the renormalization scale (RScl) parameter $C \equiv \ln(\mu^2/Q^2)$, the HT coefficient $\mu_4^{p-n}(Q^2_{\rm in})$ (at $Q^2_{\rm in}=1 \ {\rm GeV}^2$), \textcolor{black}{and the transition scale $Q_0^2$. Further, in the case of application of the $\chi$PT-motivated expression (\ref{BSRlow1}) at $Q^2 \leq Q_0^2$, there was an additional free parameter $A$ in that expression.} \textcolor{black}{The fits were performed with respect to the experimental results for BSR inelastic contributions with statistical errors.}  For the evaluation of the LT contribution of the theoretical curves at $Q^2 \geq Q_0^2$ we used the usual $\MSbar$ pQCD, and four different QCD versions with infrared-safe (and holomorphic) coupling $\A(Q^2)$: (F)APT \cite{ShS,MS96}; 2$\delta$anQCD \cite{2danQCD,anOPE,mathprg}; and a lattice-motivated 3$\delta$anQCD coupling in the three-loop and four-loop lattice MiniMOM scheme: 3l3$\delta$anQCD \cite{3l3danQCD} and 4l3$\delta$anQCD \cite{4l3danQCD}, respectively.  \textcolor{black}{At the scale $Q^2=Q_0^2$, the low-$Q^2$ and high-$Q^2$ curves were matched together.} It turned out that the three latter analytic (holomorphic) QCD versions, which agree with pQCD at large $Q^2 \gg \Lambda^2_{\rm QCD}$, give the best fit results and the \textcolor{black}{lowest values of $\chi^2$. The $\MSbar$ pQCD gives the worst results;} this is to be expected, because the $\MSbar$ pQCD coupling $a_{\rm pt}(Q^2)$ has Landau singularities at positive $Q^2 \leq 0.37 \ {\rm GeV}^2$, making the evaluation of low-$Q^2$ BSR virtually impossible. \textcolor{black}{In the low-$Q^2$ regime, the LFH QCD-motivated expression (\ref{BSRlow2}) fits better the experimental data than the $\chi$PT-motivated expression (\ref{BSRlow1}), and the resulting transition scale $Q_0^2$ is in general higher.}

\textcolor{black}{The newer experimental results \cite{data2N} from Jefferson Lab are for the squared momenta $Q^2 \geq 0.6 \ {\rm GeV}^2$ and have very small (statistical) errors. As a consequence, in the $N_f=3$ approach they represent the dominant experimental input, basically determining the theoretical curves in the regime  $Q^2 \geq 0.6 \ {\rm GeV}^2$. How well these curves (without the $\chi$PT-motivated part) describe the data below $Q^2 = 0.6 \ {\rm GeV}^2$ represents, in a way, the quality of the applied QCD approach. Our results show that the $\delta$anQCD approaches (3l3$\delta$, 4l3$\delta$, and 2$\delta$) behave in that sense better than (F)APT and $\MSbar$ pQCD approaches, cf.~Fig.~\ref{FigFitNf3} (b). This is reflected also in the obtained values of $\chi^2$ and $Q_0^2$, cf.~Tables \ref{TabResNf3}, \ref{TabResNf34}.}

As a conclusion, we can see in the example of the evaluation of BSR at low and moderate $Q^2$ that it is imperative to use QCD couplings which have no Landau singularities. While the theoretical expressions for $\Gamma_1^{p-n}(Q^2)$ with such couplings can be evaluated in principle down to $Q^2 \to 0$, this is in practice not reasonable, because these couplings are expected to be universal in the sense of being independent of the specific considered spacelike observable ${\cal D}(Q^2)$. Consequently, OPE HT terms [of the form $\sim 1/(Q^2)^n$] have to be added to the LT expression, making thus these expressions applicable only down to $Q^2 \sim 1 \ {\rm GeV}^2$. Nonetheless, as seen in the example of BSR $\Gamma_1^{p-n}(Q^2)$, these couplings allow us to evaluate such a low-momentum spacelike QCD observable to significantly lower positive values of $Q^2$ than in the usual pQCD+OPE approach; the same conclusion was drawn from the OPE application of such couplings to the evaluation of the $V$-channel Adler function ${\cal D}_V(Q^2)$ \cite{3l3danQCD,4l3danQCD}.

\textcolor{black}{We will extend \cite{WIP} the present analysis to the fits with OPE with $D=4$ term ($\sim 1/(Q^2)^2$) included, and will compare the results when the elastic contribution is excluded or included; in addition, the uncertainties of the extracted fit parameters due to the (larger) systematic errors of experimental data will be estimated.}   

\begin{acknowledgments}
\noindent
This work was supported by FONDECYT Postdoctoral Grant No.~3170116 (C.A.), by FONDECYT Regular Grant No.~1180344 (G.C. and C.A.), and by the RFBR Foundation through Grant No. 16-02-00790-a (A.V.K. and B.G.S.). C.A. thanks Bogolyubov Laboratory of Theor. Physics, of the Joint Institute for Nuclear Research, Dubna, for warm hospitality during part of this work. We thank A.L.~Kataev and J.~Bl\"umlein for important comments, \textcolor{black}{and we thank A.~Deur for bringing to our attention the newer experimental BSR data \cite{data2N}.}
\end{acknowledgments}

\appendix

\section{$\beta$ function and running coupling constant in QCD}
\label{App:A}
\def\theequation{A\arabic{equation}}
\setcounter{equation}{0}

Beta function $\beta$ takes the form of the corresponding perturbation expansion in terms of $a_{\rm pt}(Q^2)\equiv\alpha_s(Q^2)/\pi=g_s(Q^2)^2/(4\pi^2)$
\be
Q^2 \frac{d a_{\rm pt}(Q^2)}{d Q^2} = \beta(a_{\rm pt}(Q^2)), \qquad
\beta(a)=-\sum_{k=2}^\infty \beta_{k-2}a_{\rm pt}^k
\label{beta3}
\ee
with:
\bes
\label{betacoeff}
\ba
\beta_0&=&\frac{1}{4} \left( 11-\frac{2}{3}N_f \right),\nonumber\\
\beta_1&=&\frac{1}{16} \left( 102-\frac{38}{3}N_f \right),\nonumber\\
{\overline \beta}_2&=&\frac{1}{64} \left( \frac{2857}{2}-\frac{5033}{18}N_f+\frac{325}{54}N_f^2 \right),\nonumber\\
{\overline \beta}_3&=&\frac{1}{256} {\Big [} \left(\frac{149753}{6}+3564\zeta_3\right)-\left(\frac{1078361}{162}+\frac{6508}{27}\zeta_3\right)N_f
\nonumber\\
&&
+\left(\frac{50065}{162}+\frac{6472}{81}\zeta_3\right)N_f^2+\frac{1093}{729}N_f^3
{\Big ]},
\ea
\ees
where $\zeta_\nu$ is the Riemann zeta function, in particular $\zeta_3\simeq1,202057$; $N_f$ 
is the number of active quarks flavors . While the coefficients
$\beta_0$ and $\beta_1$ are universal in mass independent schemes, $\beta_k$ ($k \geq 2$) are renormalization scheme dependent. In fact, the parameters $\beta_k$ or $c_k\equiv \beta_k/\beta_0$ ($k \geq 2$) can be considered as characterizing the renormalization scheme.
In Eqs.~(\ref{betacoeff}), 
$\beta_2$ and $\beta_3$ are written in $\MSbar$ scheme.

The one-loop solution to the RGE is
\be
a_{\rm pt}^{(1-\ell)}(Q^2)=\frac{1}{\beta_0 {\rm ln}(Q^2/\Lambda^2)}, \qquad \Lambda^2=Q^2 e^{-1/(\beta_0 a(Q^2))}.
\label{1LpQCDa}
\ee
The numerical approach for the calculation
of the approximate (F)APT coupling to NLO involves the underlying pQCD coupling being the
two-loop coupling \cite{Gardi:1998qr,Magradze:1998ng,Magr} 
\ba
a_{\rm pt}^{(2-\ell)}(Q^2) = - \frac{1}{c_1} \frac{1}{\left[
1 + W_{\mp 1}(z) \right]} \ ,
\label{aptexact}
\ea
where: $c_1 = \beta_1/\beta_0$; $Q^2=|Q^2| \exp(i \phi)$; $W_{-1}$ and $W_{+1}$
are the branches of the Lambert function
for $0 \leq \phi < + \pi$ and $- \pi < \phi < 0$, 
respectively; $z$ is 
\be
z =  - \frac{1}{c_1 e} 
\left( \frac{|Q^2|}{\Lambda_{\rm L.}^2} \right)^{-\beta_0/c_1} 
\exp \left( - i {\beta_0}\phi/c_1 \right) \ ,
\label{zexpr}
\ee 
where $\Lambda_{\rm L.}$ is the Lambert QCD scale. 

In the case of 2$\delta$anQCD, the renormalization schemes of the underlying pQCD coupling $a(Q^2)$ are restricted by the requirements $M_0 \sim 1$ GeV
and $\A(0) \sim 1$. This gives: $-5.6 < c_2 < -2.0$, where
$c_2 \equiv \beta_2/\beta_0$, cf.~\cite{2danQCD,anOPE,mathprg}. 
For convenience, we can use as the central value $c_2 = -4.9$, 
and the corresponding
Lambert scheme solution of the underlying pQCD coupling $a(Q^2)$
\cite{Gardi:1998qr}
 \be
a_{\rm pt}(Q^2)=-\frac{1}{c_1}\frac{1}{1-c_2/c_1^2+W_{\mp1}(z_{\pm})} \ .
\label{aptexactc2}
\ee
In this ($c_2$-)Lambert scheme, the higher order scheme parameters
$c_k \equiv \beta_k/\beta_0$ for $k \geq 3$ are:
$c_k = c_2^{k-1}/c_1^{k-2}$.

In the case of 3$\delta$QCD in the three-loop MiniMOM scheme \cite{3l3danQCD}, the same form of the Lambert-scheme coupling is taken, with the ($N_f=3$) MiniMOM value \cite{MiniMOM} $c_2 \approx 9.3$; then $c_3=c_2^2/c_1=48.65$. In the case of 3$\delta$QCD in the four-loop MiniMOM scheme \cite{4l3danQCD}, a more complicated underlying pQCD coupling is used \cite{GCIK}, also involving Lambert functions $W_{\mp 1}(z)$ and reproducing the ($N_f=3$) four-loop MiniMOM scheme parameter values \cite{MiniMOM} $c_2 =9.297$ and $c_3=71.4538$.

\section{Charm mass contributions to BSR at NLO}
\label{App:B}
\def\theequation{B\arabic{equation}}
\setcounter{equation}{0}

The charm mass effects in BSR were calculated in Ref.~\cite{BFF}. If we ignore the $b$-quark effects (taking formally $m_b \to \infty$), the $c$-quark mass effects at NLO can be written in terms of the function $C_{\rm pBJ}^{\rm mass.,(2)}(\xi_c)$ (where $\xi_c \equiv Q^2/m_c^2$, and $m_c \approx 1.59$ GeV is the pole mass of $c$ quark) appearing in the NLO coefficient $e_2^{\rm NS}$
\bea
E_{\rm pt}(Q^2) &=& 1 - a_{\rm pt}(Q^2)_{N_f=4} + a_{\rm pt}(Q^2)^2_{N_f=4}
\left\{ -\frac{55}{12} + \frac{1}{3} \left[ N_f -1 + C_{\rm pBJ}^{\rm mass.,(2)}(\xi_c) \right] \right\} + {\cal O}(a_{\rm pt}^3),
\label{Emc1}
\eea
where $N_f=4$ and\footnote{In Ref.~\cite{BFF}, the series (\ref{Emc1}) is written in their Eq.~(6.12) where a typo in the sign before $a_{\rm pt}^2$ appeared; this is a typo, because the correct sign is used in their Eq.~(6.7).}
\bea
C_{\rm pBJ}^{\rm mass.,(2)}(\xi) & = & \frac{1}{2520} {\bigg \{} \frac{1}{\xi} ( 6 \xi^2 + 2735 \xi + 11724 ) - \frac{\sqrt{\xi+4}}{\xi^{3/2}} ( 3 \xi^3 + 106 \xi^2 + 1054 \xi + 4812) \ln \left[ \frac{ \sqrt{\xi + 4} + \sqrt{\xi} }{\sqrt{\xi + 4} - \sqrt{\xi} } \right]
\nonumber\\
&&
- 2100 \frac{1}{\xi^2} \ln^2 \left[ \frac{ \sqrt{\xi + 4} + \sqrt{\xi} }{\sqrt{\xi + 4} - \sqrt{\xi} } \right] + (3 \xi^2 + 112 \xi + 1260) \ln \xi {\bigg \}}.
\label{CBj1}
\eea
In the asymptotic limit $Q^2 \gg m_c^2$ ($\xi \gg 1$), this expression approaches slowly the value one
\be
C_{\rm pBJ}^{\rm mass.,(2)}(\xi)  = 1 - \frac{8}{3} \frac{\ln \xi}{\xi} + \frac{34}{9\xi }  + {\cal O}\left(  \frac{\ln^2 \xi}{\xi^2} \right),
\label{CBj2}
\ee
which in the limit $\xi_c \to \infty$ then reproduces in Eq.~(\ref{Emc1}) the $N_f=4$ massless expression for the NLO coefficient $e_2^{\rm NS}(N_f)$
\be
e_2^{\rm NS}(N_f)= -\frac{55}{12} + \frac{1}{3} N_f.
\label{e2NS}
\ee
We note that this convergence to the pure $N_f=4$ case (four massless quarks) is slow in BSR. For example, $3+C_{\rm pBJ}^{\rm mass.,(2)}(Q^2/m_c^2) \approx 3.13$, $3.36$, $3.73$, $3.83$ for $Q^2=5$, $10$, $50$, $100 \ {\rm GeV}^2$, respectively.

In the low-$Q^2$ regime, $Q^2 \ll m_c^2$, these corrections should reproduce the pure $N_f=3$ case (three massless quarks, $c$ quark decoupled). We can see that this is really so. Namely, the quark threshold condition \cite{CKS} relates $a_{\rm pt}(Q^2)_{N_f=4}$ and $a_{\rm pt}(Q^2)_{N_f=3}$
\be
a_{\rm pt}(Q^2)_{N_f=4} = a_{\rm pt}(Q^2)_{N_f=3} + \frac{1}{6} \ln \left( \frac{Q^2}{m_c^2} \right) a_{\rm pt}(Q^2)^2_{N_f=3} + {\cal O}(a_{\rm pt}^3),
\label{thresh}
\ee
and inserting this into the series (\ref{Emc1}) we obtain
\bea
E_{\rm pt}(Q^2) &=& 1 - a_{\rm pt}(Q^2)_{N_f=3} + a_{\rm pt}(Q^2)^2_{N_f=3}
{\bigg \{} -\frac{55}{12}       
+ \frac{1}{3} \left[ N_f  + \left( C_{\rm pBJ}^{\rm mass.,(2)}(\xi_c) - \frac{1}{2} \ln \left( \frac{Q^2}{m_c^2} \right) \right) \right] {\bigg \}} + {\cal O}(a_{\rm pt}^3),
\label{Emc2}
\eea
where $N_f=3$. The expression in parentheses, $(C_{\rm pBJ}^{\rm mass.,(2)}(\xi_c) - (1/2) \ln \xi_c))$ is regular and goes to zero when $\xi_c \equiv Q^2/m_c^2 \to 0$, as can be directly checked by expanding the expression (\ref{CBj1}) for small $\xi$
\be
C_{\rm pBJ}^{\rm mass.,(2)}(\xi)  = + \frac{1}{2} \ln \xi + \frac{2}{45} \xi \ln \xi - \frac{29}{225} \xi + {\cal O}(\xi^2 \ln \xi).
\label{CBj3}
\ee
Therefore, in $\xi_c \to 0$ limit, the $N_f=3$ QCD case is correctly reproduced. When $0 < Q^2 < 3 \ {\rm GeV}^2$, the effective flavor number $3+(C_{\rm pBJ}^{\rm mass.,(2)}(\xi_c) -  (1/2) \ln \xi_c))$ at NLO in Eq.~(\ref{Emc2}) varies between $3$ (at $Q^2 \approx 0$) and $2.85$ (at $Q^2 \approx 3 \ {\rm GeV}^2$), i.e., in the considered range of $Q^2$ the mentioned effective number of flavors is close to $N_f=3$.

\end{document}